# Three-Dimensional Optical–Electrical Simulation of $Cs_2AgBiBr_6$ Double Perovskite Solar Cells

Md Shanian Moed, Adnan Amin Siddiquee, and Md Tashfiq Bin Kashem[*]

Department of Electrical and Electronic Engineering, Ahsanullah University of Science and Technology, Dhaka – 1208, Bangladesh

[*] Corresponding author: E-mail address: tashfiq.eee@aust.edu

## Abstract

Despite significant advances in lead-free perovskite photovoltaics, achieving a balance among environmental safety, long-term stability, and high optoelectronic performance remains challenging. The inorganic double perovskite $Cs_2AgBiBr_6$ has emerged as a promising candidate owing to its robust three-dimensional crystal structure, suitable visible-range bandgap, and excellent thermal and moisture stability. However, best power conversion efficiencies (PCEs) for $Cs_2AgBiBr_6$ solar cells reported so far − 6.37% experimentally and 27.78% in numerical studies − remain below the theoretical performance potential predicted for optimized single-junction architectures, largely due to suboptimal charge transport layers, and interface-related recombination losses. Here, we address this gap using a three-dimensional (3D) finite-element method (FEM) implemented in COMSOL Multiphysics, which couples full-wave optical simulations with semiconductor drift–diffusion transport. To our knowledge, this work represents the first comprehensive 3D FEM-based study of a double halide perovskite solar cell, enabling quantitative resolution of architecture- and interface-driven losses. Screening of 25 electron transport layer (ETL)–hole transport layer (HTL) combinations identifies $CeO_2$ and P3HT as the optimal ETL and HTL respectively. Device performance is further analyzed through systematic variation of layer thicknesses, doping concentrations, defect densities, and intrinsic recombination parameters within the FTO/$CeO_2$/$Cs_2AgBiBr_6$/P3HT/Au architecture. Under optimized parameters, the simulated device achieves a PCE of 31.76%, representing the theoretical upper bound predicted by the model. Overall, this work demonstrates 3D physics-based device engineering as a decisive pathway for overcoming efficiency bottlenecks in lead-free double perovskite photovoltaics, while providing a modeling framework applicable to perovskite solar cells beyond $Cs_2AgBiBr_6$.

## 1. Introduction

The urgent need to mitigate climate change and reduce dependence on fossil fuel–based power generation has accelerated the global transition toward clean and renewable energy technologies [1], [2]. Among the available renewable resources, solar energy is the most abundant and universally accessible, making photovoltaic (PV) conversion a central pillar of future sustainable energy systems [3]. While crystalline silicon solar cells continue to dominate the commercial market due to their technological maturity, their energy-intensive manufacturing and high processing temperatures motivate continued exploration of alternative absorber materials capable of delivering high efficiency through scalable, low-temperature fabrication routes [4][5]. Third-generation photovoltaic technologies−including dye-sensitized, organic, quantum dot, and perovskite solar cells−have therefore attracted extensive research interest [6]. In particular, lead (Pb)-based hybrid organic–inorganic halide perovskites have achieved a remarkable rise in power conversion efficiency (PCE), from 3.8% in 2009 [7] to 26.95% in 2025 [8], approaching crystalline silicon benchmarks. These materials typically adopt the $ABX_3$ perovskite structure (Fig. 1a), where A denotes a larger organic or inorganic cation ($CH_3NH_3^+$, $CH(NH_2)_2^+$, $Cs^+$, $K^+$, $Na^+$ etc.), B corresponds to a divalent metal cation−most commonly $Pb^{2+}$−and X represents a halide anion ($I^-$, $Br^-$, or $Cl^-$). This crystal model enables compositional tunability while supporting the outstanding optoelectronic properties of perovskite absorbers, including strong optical absorption, long carrier diffusion lengths, low exciton binding energies, and balanced charge transport [9]. However, despite these exceptional properties, large-scale deployment remains constrained by Pb toxicity and long-term instability associated with volatile organic cations [10]–[12]. These concerns have driven the search for environmentally benign, structurally stable, lead-free alternatives.

Early substitution strategies replacing $Pb^{2+}$ with $Sn^{2+}$ or $Ge^{2+}$ were motivated by chemical similarity but suffer from rapid oxidation and severe defect formation, leading to instability and performance degradation [13]–[16]. Alternatively, $Pb^{2+}$ substitution using heterovalent cation combinations has led to the emergence of several families of lead-free halide perovskite derivatives formed to satisfy charge neutrality. These materials span a range of structural motifs, including $A_3B_2^{3+}X_9$, $A_2B^{4+}X_6$ and related low-symmetry or vacancy-ordered variants, where A-site cations are typically monovalent (e.g., $Cs^+$, $Rb^+$, $K^+$, or organic cations such as $CH_3NH_3^+/CH(NH_2)_2^+$) and B-site cations include trivalent or monovalent metals (e.g., $Bi^{3+}$, $Sb^{3+}$, $In^{3+}$, $Ag^+$, $Cu^+$, $Na^+$) [17], [18]. While these compositions incorporate environmentally benign cations, their crystal structures are typically zero- to two-dimensional in nature. This reduced dimensionality inherently limits carrier mobility and enhances recombination, resulting in suboptimal photovoltaic performance [19]–[21].

Three-dimensional halide double perovskites (HDPs) with general formula $A_2B'B''X_6$ offer a promising pathway to overcome both toxicity and dimensionality limitations [10]. In these materials, $Pb^{2+}$ is replaced by a monovalent (B′) and a trivalent (B″) cation, while preserving the three-dimensional perovskite framework (Fig. 1b). This design strategy offers substantial compositional flexibility through diverse choices of A, B′, and B″-site ions, allowing systematic tuning of band structure, carrier transport, and thermodynamic stability [22]–[25].

Among the reported HDPs, $Cs_2AgBiBr_6$ — crystallizing in the cubic elpasolite-type double perovskite structure (space group Fm$\overline{3}$m) — has attracted particular attention as a benchmark inorganic lead-free perovskite absorber [11], [26]. The cooperative incorporation of $Ag^+$ and $Bi^{3+}$ at the B′ and B″-sites enhances lattice stability, resulting in a high decomposition energy and exceptional resistance to thermal and moisture-induced degradation [27]. In addition to its robust structural stability, $Cs_2AgBiBr_6$ exhibits a suitable visible-range bandgap, long radiative and nonradiative carrier lifetimes, relatively low carrier effective masses, and notable defect tolerance—properties that are essential for efficient photovoltaic operation [10], [22], [28]. Importantly, several of these optoelectronic

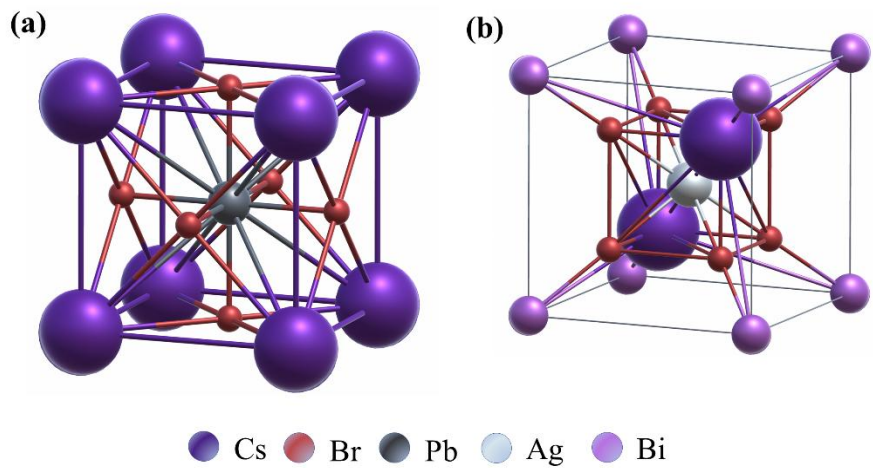

**(a)**      **(b)**

● Cs    ● Br    ● Pb    ○ Ag    ● Bi

Fig. 1. Schematic crystal structures showing the atomic site occupation of (a) perovskite and (b) double perovskite materials.

characteristics approach those of archetypal Pb-based perovskites, while offering the decisive advantages of low toxicity and long-term environmental compatibility [29]–[31].

Since its initial incorporation into photovoltaic devices by Greul *et al.* [32] using a planar FTO/TiO$_2$/Cs$_2$AgBiBr$_6$/Spiro-OMeTAD/Au architecture, which delivered a PCE of 2.43% and established the experimental feasibility of this absorber, Cs$_2$AgBiBr$_6$ has been explored extensively across diverse device architectures, contact layer combinations, and deposition routes, including both solution-processed and vapor-deposited films [33], [34]. Continuous experimental progress has led to the current record efficiency of 6.37%, achieved by Zhang *et al.* via optimized planar architectures (ITO/SnO$_2$/Cs$_2$AgBiBr$_6$/Spiro-OMeTAD/Au), demonstrating excellent operational stability (1440h at 85°C) as well [28]; however, the performance remains modest relative to Pb-based counterparts. In parallel, numerical studies suggest that significantly higher efficiencies may be achievable through appropriate selection of charge transport layers and careful optimization of material and interfacial parameters [35]–[38]. For instance, Danladi et al. reported a PCE of 25.56% for the FTO/ZnO/Cs$_2$AgBiBr$_6$/CFTS/Au structure [36]. Hakami *et al.* proposed an all-inorganic device architecture based on FTO/TiO$_2$/Cs$_2$AgBiBr$_6$/NiO/Au, achieving a simulated PCE of 26.71% [37]. Similarly, Srivastava *et al.* identified an optimized planar FTO/AZO/Cs$_2$AgBiBr$_6$/ZnTe configuration, predicting a PCE of 26.89% [35]. More recently, Raj *et al.* reported an optimized efficiency of 27.78% for the FTO/TiO$_2$/Cs$_2$AgBiBr$_6$/Cu$_2$O/Au structure [38]. Collectively, these insights motivate a systematic device-level optimization strategy in which suitable electron transport layer (ETL)–hole transport layer (HTL) combinations are first identified, followed by targeted tuning of layer thicknesses, doping concentrations, and bulk and interfacial defect densities, with the aim of identifying realistic pathways toward performance enhancement in Cs$_2$AgBiBr$_6$-based solar cells. Despite the advances, most prior modeling efforts rely on one-dimensional drift–diffusion solvers that assume simplified generation profiles and neglect spatial optical redistribution and multidimensional interfacial effects. Therefore, realistic modeling of optical field distribution and its coupling to carrier transport is essential for identifying true performance limits and dominant loss mechanisms.

In this context, physics-based finite-element modeling offers a powerful framework for resolving coupled optical and electrical phenomena in heterogeneous device stacks. COMSOL Multiphysics enables self-consistent solutions of Maxwell's equations and semiconductor transport equations within multidimensional geometries, allowing spatially resolved analysis of photogeneration, band bending, quasi-Fermi level splitting, and recombination processes [39]–[42]. However, such a coupled three-dimensional (3D) treatment remains largely unexplored for Cs$_2$AgBiBr$_6$-based solar cells.

Motivated by this gap, the present work develops a comprehensive 3D finite-element opto-electrical framework to investigate Cs$_2$AgBiBr$_6$ solar cells across multiple charge transport layer combinations. Twenty-five ETL–HTL pairings are systematically screened to identify optimal contact materials. Building upon the most favorable configuration, layer thicknesses, doping concentrations, bulk and interfacial defect densities, intrinsic recombination coefficients, temperature, and illumination intensity are rigorously examined. By explicitly calculating optical generation profiles and coupling them to drift–diffusion transport, this integrated framework establishes physically grounded performance bounds, identifies critical design tolerance windows, and explains the fundamental loss mechanisms governing Cs$_2$AgBiBr$_6$ photovoltaics, thereby providing quantitative guidance toward experimentally realizable, high-efficiency lead-free double perovskite solar cells.

## 2. Simulation Framework

### 2.1 Device Structure and Material Parameters

All simulations were carried out using a three-dimensional finite-element method in COMSOL Multiphysics framework. The investigated device architecture adopts a layered heterojunction configuration of FTO/ETL/Cs$_2$AgBiBr$_6$/HTL/Au (Fig. 2a). A schematic of the equilibrium energy-band alignment of the complete device stack is shown in Fig. 2b.

Fluorine-doped tin oxide (FTO) is employed as the transparent front electrode due to its high optical transmittance in the visible region and excellent chemical and thermal stability, making it a well-established and reliable choice for perovskite and inorganic thin-film photovoltaic devices [43].

An ETL is introduced between FTO and the absorber to enable selective extraction of photogenerated electrons while suppressing hole transport toward the front contact. To assess the influence of transport-layer properties, five representative ETL materials − AZO (aluminum-doped zinc oxide), CeO$_2$ (cerium (IV) oxide), WS$_2$ (tungsten disulphide), C60 (buckminsterfullerene) and PCBM ([6,6]-phenyl-C61-butyric acid methyl ester) − were examined. These materials span a broad range of electronic characteristics, including wide-bandgap metal oxides, layered inorganic semiconductors, and organic fullerene derivatives [44][45]. Such diversity allows a comprehensive evaluation of band alignment, carrier mobility, optical transparency, and interfacial recombination behavior within the device.

Cs$_2$AgBiBr$_6$ is employed as the absorber layer, where photon absorption and electron–hole pair generation occur under illumination. In the simulated architecture, carrier transport within the absorber is governed by bulk material properties, defect-assisted recombination, and interfacial charge transfer at the ETL and HTL interfaces.

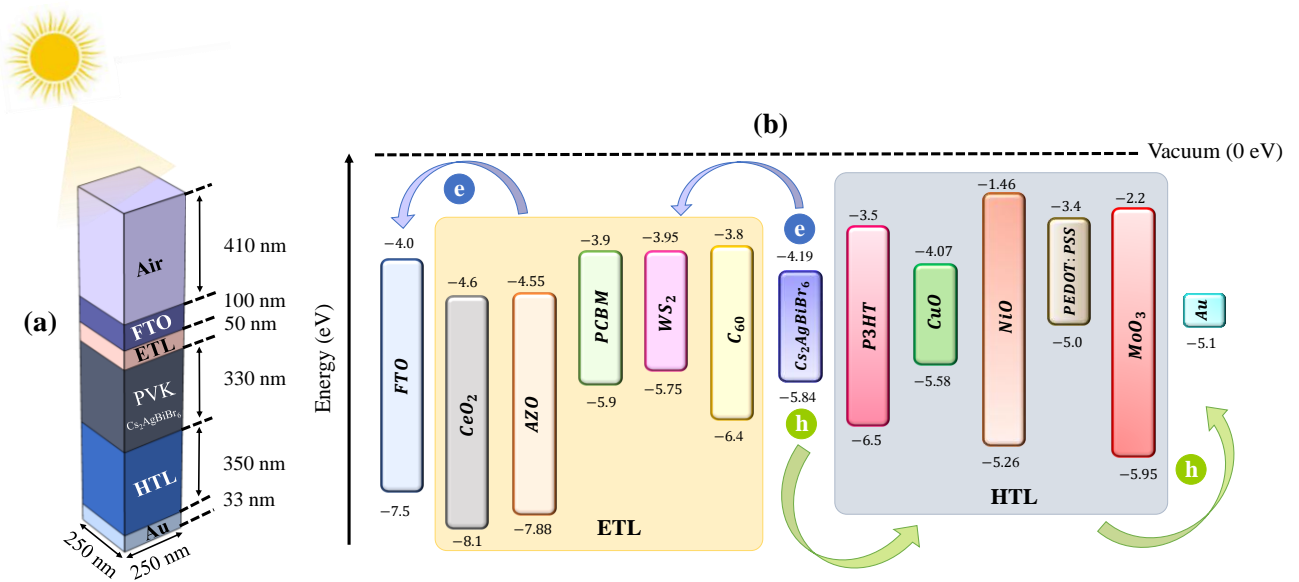

Fig. 2. (a) Schematic illustration of the Cs$_2$AgBiBr$_6$ perovskite (PVK) absorber-based solar cell architecture, with the individual layer thicknesses; detailed material parameters are listed in Tables 1-3. (b) Energy band alignment diagram showing the conduction band minima (CBM) and valence band maxima (VBM) of the constituent layers; for each material, the upper and lower numerical values denote the CBM and VBM energy levels, respectively. Five electron transport layers (ETLs)—CeO$_2$, AZO, PCBM, WS$_2$, and C60—and five hole transport layers (HTLs)—P3HT, CuO, NiO, PEDOT:PSS, and MoO$_3$—were investigated. The conduction and valence band offsets (CBO, VBO) at the absorber/ETL and absorber/HTL interfaces are summarized in Tables 4 and 5. The symbols "e" and "h" in panel (b) represent electrons and holes. Under illumination, photogenerated electron–hole pairs are created within the Cs$_2$AgBiBr$_6$ absorber; electrons are transported toward the FTO front contact through the ETL, while holes are collected at the Au back contact via the HTL.

On the rear side of the absorber, a HTL is incorporated to facilitate efficient extraction of photogenerated holes while blocking electron transport toward the back contact. Five HTL materials − P3HT (poly(3-hexylthiophene)), PEDOT:PSS (Poly(3,4-ethylenedioxythiophene) polystyrene sulfonate), CuO (copper (II) oxide), NiO (nickel (II) oxide) and $MoO_3$ (molybdenum trioxide) − were considered, representing polymeric, oxide, and transition-metal-oxides, with distinct valence-band alignments and hole transport characteristics. These materials have been extensively reported as effective HTLs in perovskite and thin-film solar cells and provide a suitable basis for systematic transport-layer engineering [46], [47].

Gold (Au), with a high work function of approximately 5.1 eV, is selected as the back contact to ensure favorable energetic alignment with the HTL and efficient hole collection while minimizing contact resistance. The top and bottom electrodes of the device were modeled as Ohmic contacts by applying Dirichlet boundary conditions for the electrostatic potential. In COMSOL, these conditions directly fix the potential at the contact nodes, ensuring that the applied voltage is correctly enforced during the solution of the semiconductor equations. The applied voltage bias was introduced at the back contact, while the front contact was grounded. All other external boundaries were treated as electrically insulating to prevent unintended current flow. An air domain was included above the FTO front electrode to represent the surrounding optical medium and to correctly impose the incident electromagnetic boundary conditions. All the material parameters − including bandgaps, electron affinities, dielectric constants, and carrier mobilities − were adopted from previously reported literature.

Table 1. Input parameters of FTO, $Cs_2AgBiBr_6$, and HTL materials: band gap ($E_g$), electron affinity ($\chi$), relative dielectric permittivity ($\varepsilon_r$), effective density of states in the conduction band ($N_C$), effective density of states in the valence band ($N_V$), electron mobility ($\mu_n$), hole mobility ($\mu_p$), acceptor density ($N_A$), donor density ($N_D$), and defect density ($N_t$)

| Parameters | FTO | $Cs_2AgBiBr_6$ | P3HT | CuO | NiO | PEDOT:PSS | $MoO_3$ |
|---|---|---|---|---|---|---|---|
| Thickness (nm) | 100 | 300 | 350 | 350 | 350 | 350 | 350 |
| $E_g$ (eV) | 3.5 | 1.654 | 1.7 | 1.51 | 3.8 | 1.6 | 3.75 |
| $\chi$ (eV) | 4 | 4.19 | 3.5 | 4.07 | 1.46 | 3.4 | 2.2 |
| $\varepsilon_r$ | 9 | 5.8 | 3 | 18.1 | 10.7 | 3 | 4.45 |
| $N_c$ (1/cm$^3$) | $2.2 \times 10^{18}$ | $1 \times 10^{16}$ | $2 \times 10^{21}$ | $2.2 \times 10^{19}$ | $2.8 \times 10^{19}$ | $2.2 \times 10^{18}$ | $1 \times 10^{19}$ |
| $N_v$ (1/cm$^3$) | $1.8 \times 10^{19}$ | $1 \times 10^{16}$ | $2 \times 10^{21}$ | $5.5 \times 10^{20}$ | $10^{19}$ | $1.8 \times 10^{19}$ | $1 \times 10^{19}$ |
| $\mu_n$ (cm$^2$/Vs) | $2 \times 10^1$ | 1.81 | $1.8 \times 10^{-3}$ | 100 | 12 | $4.5 \times 10^{-2}$ | $1.1 \times 10^3$ |
| $\mu_p$ (cm$^2$/Vs) | $1 \times 10^1$ | 0.49 | $1.86 \times 10^{-2}$ | 0.1 | 2.8 | $4.5 \times 10^{-2}$ | $1.1 \times 10^3$ |
| $N_A$ (1/cm$^3$) | 1 | $3 \times 10^{15}$ | $10^{18}$ | $1 \times 10^{18}$ | $1 \times 10^{18}$ | $1 \times 10^{18}$ | $2 \times 10^{18}$ |
| $N_D$ (1/cm$^3$) | $1 \times 10^{19}$ | 0 | 0 | 0 | 0 | 0 | 0 |
| $N_t$ (1/cm$^3$) | $10^{14}$ | $10^{14}$ | $10^{14}$ | $10^{15}$ | $10^{15}$ | $10^{15}$ | $10^{15}$ |
| Reference | [37] | [48] | [49] | [50] | [51] | [52] | [53] |

Table 2. Input parameters of ETL materials

| Parameters | AZO | CeO$_2$ | PCBM | WS$_2$ | C$_{60}$ |
|---|---|---|---|---|---|
| Thickness (nm) | 30 | 30 | 30 | 30 | 30 |
| $E_g$ (eV) | 3.33 | 3.5 | 2 | 1.8 | 1.7 |
| $\chi$ (eV) | 4.55 | 4.6 | 3.9 | 3.95 | 3.9 |
| $\varepsilon_r$ | 8.12 | 9 | 3.9 | 13.6 | 4.2 |
| $N_C$ (1/cm$^3$) | $4.1 \times 10^{18}$ | $1 \times 10^{20}$ | $2 \times 10^{21}$ | $1 \times 10^{18}$ | $8 \times 10^{19}$ |
| $N_V$ (1/cm$^3$) | $8.2 \times 10^{19}$ | $2 \times 10^{21}$ | $2 \times 10^{21}$ | $2.4 \times 10^{19}$ | $8 \times 10^{19}$ |
| $\mu_n$ (cm$^2$/Vs) | 20 | 100 | 0.2 | 100 | $8 \times 10^{-2}$ |
| $\mu_p$ (cm$^2$/Vs) | 10 | 25 | 0.2 | 100 | $8 \times 10^{-3}$ |
| $N_A$ (1/cm$^3$) | 0 | 0 | 0 | 0 | 0 |
| $N_D$ (1/cm$^3$) | $1 \times 10^{19}$ | $2 \times 10^{21}$ | $2.93 \times 10^{17}$ | $10^{18}$ | $1 \times 10^{17}$ |
| $N_t$ (1/cm$^3$) | $10^{15}$ | $10^{15}$ | $10^{15}$ | $10^{15}$ | $10^{15}$ |
| Reference | [54] | [50] | [55] | [56] | [55] |

Table 3. Input parameters for interface and bulk defects

| Parameters | ETL/Cs$_2$AgBiBr$_6$ interface | Cs$_2$AgBiBr$_6$/HTL interface | Bulk Cs$_2$AgBiBr$_6$ |
|---|---|---|---|
| Defect type | Neutral | Neutral | Neutral |
| Electron capture cross section (cm$^2$) | $10^{-18}$ | $10^{-18}$ | $10^{-15}$ |
| Hole capture cross section (cm$^2$) | $10^{-18}$ | $10^{-18}$ | $10^{-15}$ |
| Energetic distribution | Single | Single | Single |
| Thermal velocity (cm/s) | $1 \times 10^7$ | $1 \times 10^7$ | $1 \times 10^7$ |
| Impurity energy level | Midgap | Midgap | Midgap |
| Total density (cm$^{-2}$) | $10^{10}$ (Variable) | $10^{10}$ (Variable) | $10^{14}$ cm$^{-3}$ (Variable) |

Table 4. CBO between Cs$_2$AgBiBr$_6$ and ETL materials: CBO = $\chi_{absorber} - \chi_{ETL}$

| ETL | AZO | CeO$_2$ | PCBM | WS$_2$ | C$_{60}$ |
|---|---|---|---|---|---|
| CBO | 0.35 | 0.4 | -0.3 | -0.25 | -0.3 |

Table 5. VBO between Cs$_2$AgBiBr$_6$ and HTL materials: VBO = $(\chi_{HTL} + E_{g,HTL}) - (\chi_{absorber} + E_{g,absorber})$

| HTL | P3HT | CuO | NiO | PEDOT-PSS | MoO$_3$ |
|---|---|---|---|---|---|
| VBO | 0.04 | -0.34 | -0.02 | 0.24 | -0.71 |

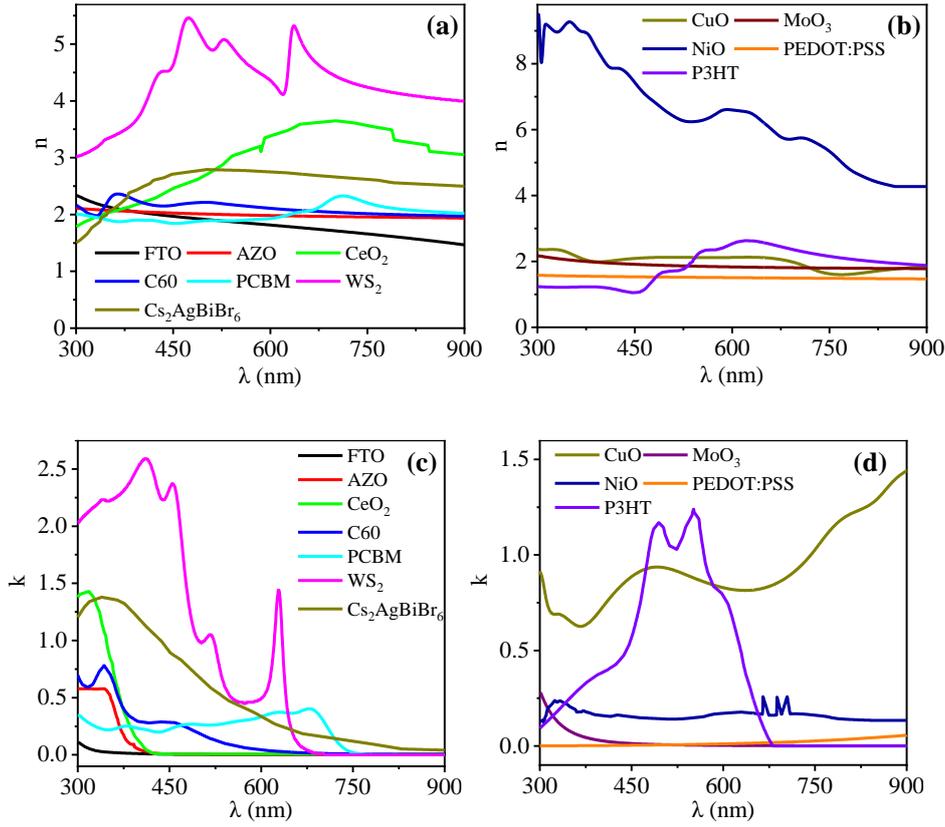

Fig. 3. (a,b) Refractive index (n) and (c,d) extinction coefficient (k) spectra of the FTO, absorber, ETL and HTL materials used in the simulations, extracted from Refs. [48], [57]–[67].

Unless otherwise stated, all simulations were performed under standard AM1.5G illumination with an incident power density of 1000 Wm$^{-2}$ at an operating temperature of 300 K. The complete set of material, and interface parameters employed in the simulations is summarized in Tables 1-5 and Fig. 3.

## 2.2 Mesh Generation

A user-controlled, structured mesh was employed for the six-layer device architecture. Each domain was discretized individually with minimum and maximum element sizes of 1.91 nm and 22 nm, respectively. The maximum element growth rate was set to 1.08 with a curvature factor of 0.3 and narrow region resolution of 0.95 to ensure accurate resolution of sharp interfacial gradients. A hybrid meshing scheme combining free triangular elements (surface) and swept quadrilateral elements (volume) was implemented. The top and bottom faces were meshed using free triangular elements, while remaining faces were discretized using a swept quadrilateral sequence. Thickness-dependent element distributions were applied across domains using a combination of linear and exponential growth profiles. For example, the 50 nm ETL layer was discretized into 80 elements with symmetric linear grading, whereas the perovskite absorber was initially resolved with 50 elements using exponential grading (growth ratio = 5), ensuring refined meshing near the ETL/perovskite heterointerface. Tessellation was set to automatic. Adaptive mesh refinement was enabled to dynamically resolve regions exhibiting large gradients in electrostatic potential, carrier concentration, and recombination rate. Depending on the device configuration, the total number of degrees of freedom (DOFs) solved ranged from approximately $3.24 \times 10^4$ for the initial structure to $2.58 \times 10^5$ for the optimized device. Mesh independence was confirmed by progressively increasing mesh density until the current–voltage characteristics exhibited negligible variation. Mesh quality was evaluated using skewness and growth rate metrics; the resulting distributions satisfied standard quality criteria, confirming numerical stability and solution convergence.

## 2.3 Theoretical Formulation

The simulations were performed by coupling the Electromagnetic Waves, Frequency Domain (EWFD) interface with the Semiconductor interface within COMSOL Multiphysics. The coupled approach enables the calculation of optical absorption and photogeneration using full-wave electromagnetic analysis, followed by charge carrier transport using a drift–diffusion formalism.

### 2.3.1 Optical Simulation Using EWFD Interface

The optical response of the device under illumination was computed using the EWFD interface, which solves Maxwell's equations in three dimensions in the frequency domain to obtain the wavelength-dependent electric field distribution, $\mathbf{E}(\lambda)$ [68]:

$$\boldsymbol{\nabla} \times \mu_r^{-1}(\boldsymbol{\nabla} \times \boldsymbol{E}) - k_0^2 \varepsilon_r \boldsymbol{E} = 0 \tag{1}$$

where $\mathbf{E}$ denotes the electric field vector, $\varepsilon_r$ and $\mu_r$ are the relative permittivity and permeability of the materials, respectively, and $k_0$ is the free-space wave number. $\varepsilon_r$ and $k_0$ are expressed as [69],

$$\varepsilon_r = \varepsilon' + \varepsilon'' = [n(\lambda) - jk(\lambda)]^2 \tag{2}$$

$$k_0 = \frac{\omega}{c_0} \tag{3}$$

where $\varepsilon'$ and $\varepsilon''$ are the real and imaginary parts of $\varepsilon_r$, respectively; n and k are refractive index and extinction coefficient, respectively; $\omega$ is the angular frequency of the incident radiation, and $c_0$ is the speed of light in vacuum.

A normally incident plane wave corresponding to the AM1.5G solar spectrum was applied at the FTO side to represent solar irradiation. The wavelength-dependent n and k of each layer were incorporated to account for optical dispersion and absorption. The local volumetric absorbed power density at each wavelength was calculated as:

$$Q_{abs}(x, y, z, \lambda) = \frac{1}{2} \omega(\lambda) \varepsilon_0 \varepsilon''(\lambda) |\boldsymbol{E}(x, y, z, \lambda)|^2 \tag{4}$$

Here, $\varepsilon_0$ is the vacuum permittivity, and |E| represents the magnitude of the electric field. The photogeneration rate at each wavelength was then computed as [69],

$$G_{opt}(x, y, z, \lambda) = \frac{Q_{abs}(x,y,z,\lambda)}{h\nu(\lambda)} \tag{5}$$

Here h is Planck's constant and $\nu$ is the photon frequency. To account for the full spectral range of illumination, the wavelength-dependent generation rate was integrated over the relevant wavelength interval for each layer:

$$G_{tot}(x, y, z) = \int_{\lambda_{min}}^{\lambda_{max}} G_{opt}(\lambda) d\lambda \qquad (6)$$

The spectrally integrated generation profile was then volume-averaged and applied to the corresponding layers in Semiconductor physics module, where it served as the photogeneration term in the carrier continuity equations for drift–diffusion simulations.

### 2.3.2 Charge Transport Modeling in Semiconductor Interface

Electrical modeling was performed using the Semiconductor interface, which solves Poisson's equation coupled with the electron and hole continuity equations within the drift–diffusion approximation.

The electrostatic potential $\Phi$ is governed by Poisson's equation:

$$\boldsymbol{\nabla}.(\varepsilon\boldsymbol{\nabla}\Phi) = -q(p - n + N_D^+ - N_A^-) \qquad (7)$$

where $\varepsilon$ is the relative permittivity, q is the elementary charge, n and p are the electron and hole concentrations, and $N_D^+$ and $N_A^-$ represent ionized donor and acceptor densities.

The continuity equations for electrons and holes are given by:

$$\boldsymbol{\nabla}.\boldsymbol{J}_n = q(G_n - R_n) \qquad (8)$$

$$\boldsymbol{\nabla}.\boldsymbol{J}_p = q(G_p - R_p) \qquad (9)$$

where G is the total carrier generation rate, including photogeneration, and R is the recombination rate. The subscripts n and p correspond to electron and hole, respectively.

The electron and hole current densities are expressed as:

$$\boldsymbol{J}_n = q\mu_n n\boldsymbol{\nabla}\Phi + qD_n\nabla n$$
$$\qquad (10)$$

$$\boldsymbol{J}_p = q\mu_p p\boldsymbol{\nabla}\Phi - qD_p\nabla p \qquad (11)$$

where $\mu$ is the carrier mobility, and D is the diffusion coefficient, related to the mobilities through the Einstein relation:

$$D_{n,p} = \mu_{n,p} \frac{kT}{q} \qquad (12)$$

Where $k$ is the Boltzmann constant and T is temperature. Carrier recombination within the device was modeled using three primary mechanisms: Shockley–Read–Hall (SRH), radiative (direct), and Auger recombination. SRH recombination accounts for defect-mediated, nonradiative recombination:

$$R_{SRH} = \frac{np - n_i^2}{\tau_p(n + n_i) + \tau_n(p + n_i)} \qquad (13)$$

where $n_i$ is the intrinsic carrier concentration, and $\tau_n$ and $\tau_p$ are the electron and hole lifetimes, respectively. Radiative (direct) recombination describes band-to-band recombination:

$$R_{Rad} = C_{Radiative}(np - n_i^2) \tag{14}$$

where $C_{Radiative}$ is the radiative recombination coefficient. Auger recombination models three-particle interactions, significant at high carrier densities:

$$R_{Auger} = C_{n,Auger}\, n(np - n_i^2) + C_{p,Auger}\, p(np - n_i^2) \tag{15}$$

where $C_{n,Auger}$ and $C_{p,Auger}$ are the Auger coefficients for electrons and holes, respectively.

### 2.3.3 Performance Parameters

The coupled optical–electrical problem was solved using a stationary solver. From the converged solutions, current density–voltage (J−V) characteristics under illumination were extracted, and key photovoltaic parameters—including short-circuit current density ($J_{SC}$), open-circuit voltage ($V_{OC}$), fill factor (FF), and PCE—were calculated.

$$V_{OC} = \frac{kT}{q} ln\left(\frac{J_{SC}}{J_0} + 1\right) \tag{16}$$

$$FF = \frac{V_m J_m}{V_{OC} J_{SC}} \tag{17}$$

$$PCE = \frac{V_{OC} J_{SC} FF}{P_{in}} \tag{18}$$

Where $J_0$ is the reverse saturation current density, $V_m$ and $J_m$ refer to the voltage and current density respectively at the maximum power point, and $P_{in}$ is the input power.

## 3. Results and Analysis

### 3.1 Model Validation and Benchmarking

To establish the reliability of the proposed simulation framework, the device model was first validated by reproducing experimentally reported results of a previously published $Cs_2AgBiBr_6$-based solar cell with the configuration: ITO/$SnO_2$/$Cs_2AgBiBr_6$/Spiro-OMeTAD/Au [28]. The simulation parameters, including layer thicknesses and material properties were carefully adopted from the experimental study to ensure a meaningful comparison.

Figure 4 presents a direct comparison between the simulation and experimental results in terms of J−V characteristics under AM1.5G illumination. The simulated curve closely reproduces the experimental response across the entire bias range, capturing both the slope near open-circuit conditions and the curvature around the maximum power point. This agreement indicates that the model accurately describes the dominant carrier transport and recombination processes governing device operation.

In addition to the J-V characteristics, the extracted photovoltaic performance parameters—$V_{OC}$, $J_{SC}$, FF, and PCE—are summarized in the Table inside the figure and compared with the values reported in Ref. [28]. The simulated results show close quantitative agreement with the measured data, with only minor deviations that can reasonably be attributed to experimental uncertainties, material inhomogeneities, and processing-induced variations not explicitly captured in the numerical model.

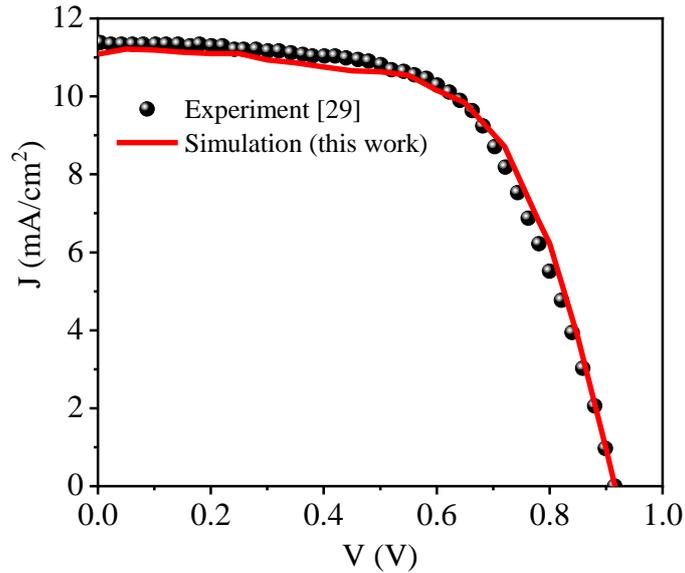

Fig. 4. Benchmarking and validation of the simulation framework presented in this work through comparison with experimentally reported J–V characteristics and associated photovoltaic parameters ($V_{OC}$, $J_{SC}$, FF, and PCE) of $Cs_2AgBiBr_6$-based solar cell [28].

Overall, the strong agreement between simulation and experiment confirms the validity of the adopted physical models and parameter sets. This validation provides a solid foundation for systematic transport-layer engineering, followed by broader material and device-level parametric investigations aimed at optimizing the performance of the $Cs_2AgBiBr_6$-absorber-based solar cell architecture.

### 3.2 Performance Analysis of ETL-HTL Pairings

A comprehensive charge transport-layer (ETL and HTL) study with 25 distinct ETL/HTL combinations was carried out to identify optimal carrier-selective contacts for the $Cs_2AgBiBr_6$ absorber. The selected transport materials satisfy established selection criteria reported in prior literature [70], including suitable band alignment, sufficient carrier mobility, optical transparency, and chemical stability. $V_{OC}$, $J_{SC}$, FF, and PCE were extracted for each combination (Fig. 5), a pronounced dependence of device performance on transport-layer pairing is observed, underscoring the critical role of interfacial energetics and carrier selectivity.

Higher $V_{OC}$ exceeding ~1.8 V are achieved for AZO/P3HT, AZO/NiO, $CeO_2$/PEDOT:PSS, $WS_2$/CuO, and $WS_2$/$MoO_3$ configurations (Fig. 5a). These higher voltages can be attributed to favorable interfacial band alignment and effective suppression of interfacial recombination, which enhance quasi-Fermi level splitting within the absorber. In contrast, combinations with suboptimal energy-level matching exhibit lower $V_{OC}$, highlighting the sensitivity of voltage output to interface energetics.

Highest $J_{SC}$ of 16.45 mA $cm^{-2}$ is obtained for $CeO_2$/P3HT combination (Fig. 5b). This enhanced current response reflects the excellent optical transparency of $CeO_2$ in the visible spectrum, combined with efficient electron extraction and minimal parasitic absorption in the ETL. Simultaneously, the favorable hole mobility and interfacial compatibility of P3HT facilitate efficient charge collection, leading to superior photogenerated current compared with alternative transport-layer pairings.

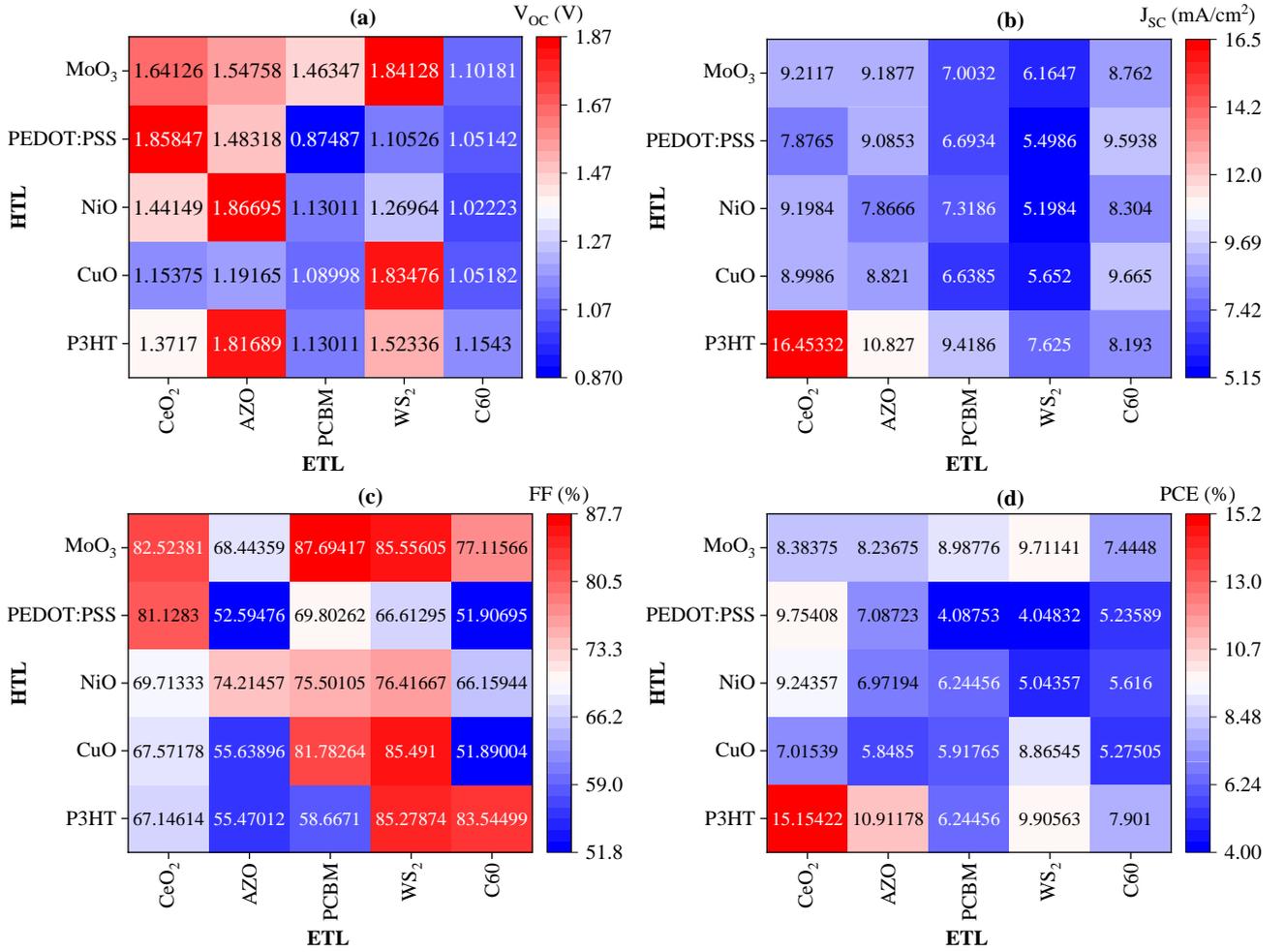

Fig. 5. Comparison of performance parameters for 25 ETL/HTL combinations (5 ETLs × 5 HTLs), showing (a) $V_{OC}$, (b) $J_{SC}$, (c) FF, and (d) PCE. All the simulations were performed using the following layer thicknesses along the z-axis: FTO = 100 nm, ETL = 50 nm, $Cs_2AgBiBr_6$ = 330 nm, HTL = 350 nm, and Au = 33 nm.

FF also varies considerably with transport-layer pairing (Fig. 5c). High FF exceeding 85% are achieved for PCBM/$MoO_3$, $WS_2$/$MoO_3$, $WS_2$/CuO, and $WS_2$/P3HT configurations. These combinations benefit from reduced series resistance, efficient carrier selectivity, and suppressed recombination losses near the maximum power point, resulting in improved diode rectification and more ideal J–V characteristics.

As a cumulative metric, the PCE reflects the combined trends in $V_{OC}$, $J_{SC}$, and FF (Fig. 5d). The highest PCE of 15.15% is obtained for the $CeO_2$/P3HT configuration, which simultaneously offers the largest $J_{SC}$ along with competitive $V_{OC}$ and FF values. Other combinations exhibiting high voltage or fill factor alone do not achieve comparable efficiencies due to trade-offs in current generation or resistive losses. These results indicate that balanced optimization of charge extraction, optical transmission, and interfacial recombination is essential for maximizing device efficiency.

Based on this comprehensive comparative assessment, $CeO_2$ and P3HT emerge as the most promising ETL and HTL, respectively, for $Cs_2AgBiBr_6$ absorber-based solar cell. Accordingly, this ETL/HTL pair is selected for subsequent wavelength-resolved optical analysis and detailed material and device-level parametric optimization.

## 3.3 Optical Response and Carrier Generation Analysis

### 3.3.1 Wavelength-Dependent Optical Field, Absorbed Power, and Carrier Generation

In this section, the spatial distributions of the optical electric field, local volumetric absorbed power density ($Q_{abs}$), and carrier generation rate within the full device stack with the optimal charge transport layers ($CeO_2$ and P3HT), obtained from EWFD module (Fig. 6-8), have been analyzed. 3D visualizations provide a global perspective on light propagation, field localization, and absorption across all layers (Fig. 6-8, a-e). To enable clearer layer-resolved interpretation, two-dimensional (2D) cross-sectional maps were subsequently extracted in the yz-plane at a fixed lateral position (x =125 nm), which intersects all functional layers from the FTO front contact to the Au back electrode (Fig. 6-8, f-j).

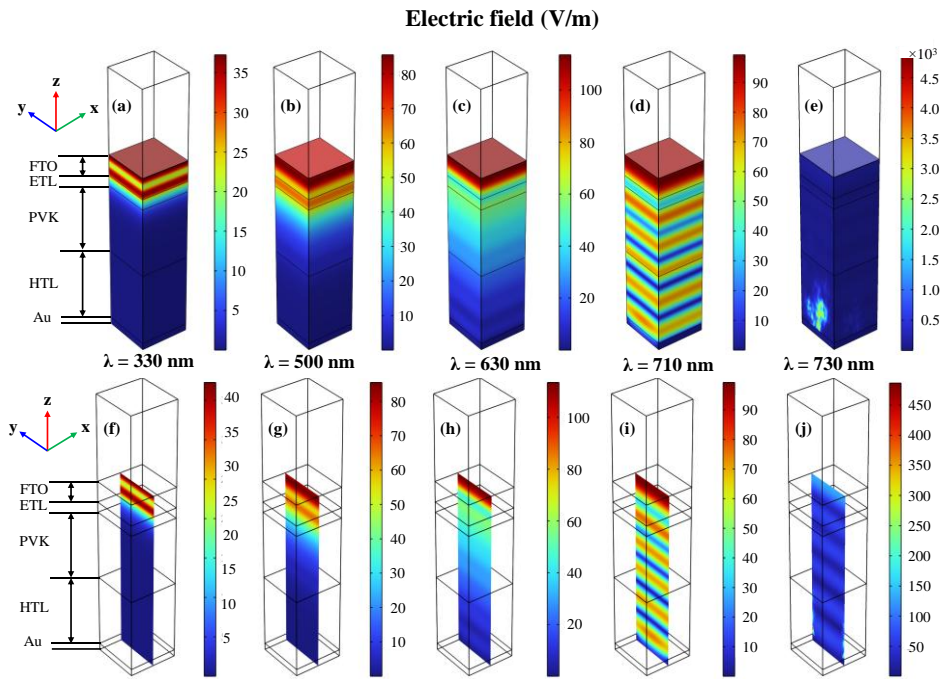

Fig. 6. (a−e) 3D and (f−j) corresponding two-dimensional (2D) cross-sectional distributions of the optical electric field at representative wavelengths. The 2D cross sections are taken in the yz plane at a fixed lateral position of x = 125 nm. Panels (a,f), (b,g), (c,h), (d,i), and (e,j) correspond to λ = 330, 500, 630, 710, and 730 nm, respectively. The color scales represent the electric-field magnitude in V/m. PVK corresponds to the absorber layer.

**Power density (W/cm³)**

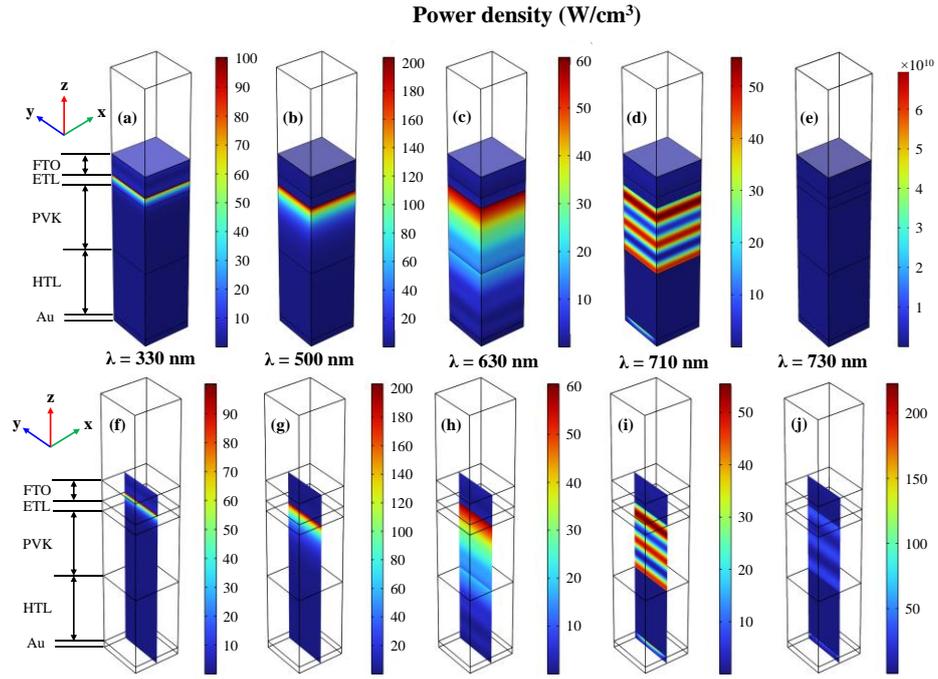

Fig. 7. (a–e) 3D and (f–j) corresponding 2D cross-sectional distributions of the local absorbed power density at representative wavelengths. The 2D cross sections are taken in the yz plane at a fixed lateral position of x = 125 nm. Panels (a,f), (b,g), (c,h), (d,i), and (e,j) correspond to λ = 330, 500, 630, 710, and 730 nm, respectively. The color scales represent the absorbed power density in W/cm³. PVK corresponds to the absorber layer.

**Carrier generation rate (cm⁻³s⁻¹)**

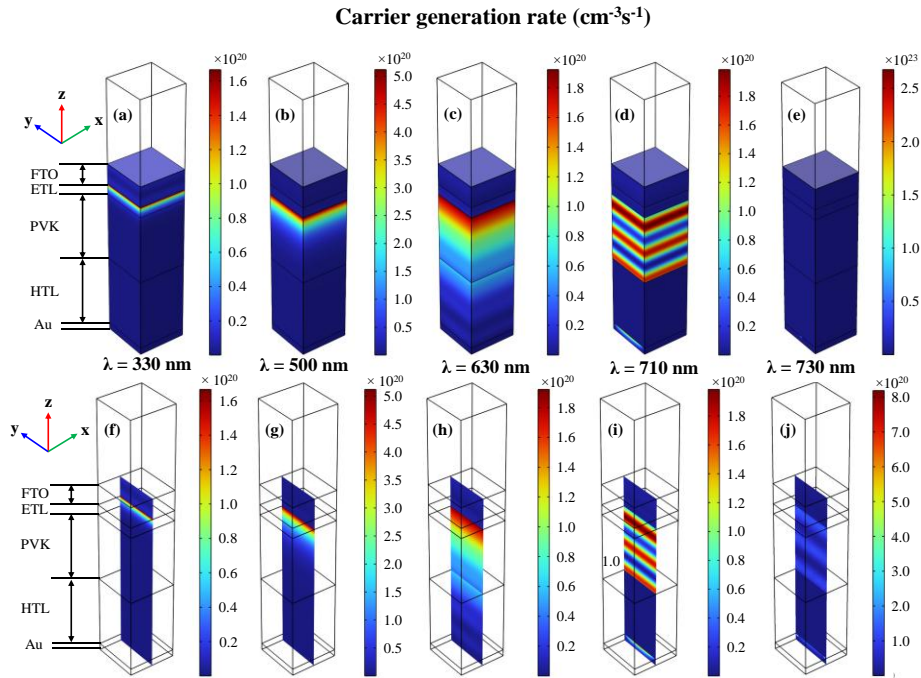

Fig. 8. (a–e) 3D and (f–j) corresponding 2D cross-sectional distributions of the carrier (electron-hole pair) generation rate at representative wavelengths. The 2D cross sections are taken in the yz plane at a fixed lateral position of x = 125 nm. Panels (a,f), (b,g), (c,h), (d,i), and (e,j) correspond to λ = 330, 500, 630, 710, and 730 nm, respectively. The color scales represent the generation rate in cm⁻³s⁻¹. PVK corresponds to the absorber layer.

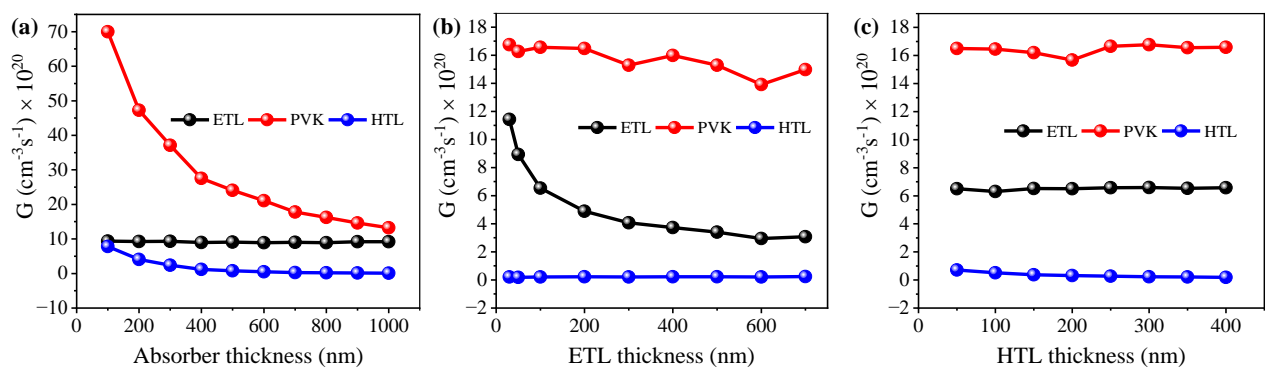

Fig. 9. Wavelength-integrated volume-averaged carrier generation rate inside the absorber (PVK in the legend), ETL, and HTL as a function of layer thickness: (a) absorber, (b) ETL, and (c) HTL. For (a), ETL and HTL thicknesses were fixed at 50 nm and 350 nm, respectively; for (b), absorber and HTL thicknesses were fixed at 800 nm and 350 nm; and for (c), absorber and ETL thicknesses were fixed at 800 nm and 100 nm.

### 3.3.2 Thickness-Dependent Carrier Generation in ETL, Absorber, and HTL

Following the wavelength-resolved optical analysis, the influence of individual layer thickness on the volume-averaged carrier generation rate was next examined (Fig. 9). The thicknesses of the absorber, ETL, and HTL were varied independently while keeping the remaining layers fixed, allowing the role of each layer in governing overall photogeneration to be isolated.

When the thickness of the $Cs_2AgBiBr_6$ absorber was increased, a monotonic decrease in the generation rate within the absorber was observed (Fig. 9a). This trend arises because optical absorption in $Cs_2AgBiBr_6$ is strongly front-loaded, with most photons absorbed near the ETL/absorber interface, as established by the earlier $Q_{abs}$ and generation profiles. Increasing absorber thickness therefore enlarges the optically weakly excited rear region, reducing the average generation rate when normalized over the full absorber volume. In contrast, the generation rates within the ETL and HTL remain nearly unchanged and negligibly small, confirming that photogeneration is predominantly confined to the absorber layer and is insensitive to absorber thickness variations in the adjacent transport layers.

A similar but weaker trend is observed when varying the ETL thickness (Fig. 9b). Increasing the $CeO_2$ thickness from 30 nm to 700 nm leads to a slight reduction in the generation rate within the absorber, accompanied by a continuous decrease in generation within the ETL itself. The latter reflects the dilution of absorbed power over a thicker wide-bandgap layer, while the former originates from increased optical attenuation and parasitic absorption in the ETL, which reduce photon flux reaching the absorber. Throughout this thickness range, the generation rate within the HTL remains essentially zero, signifying negligible absorption in the investigated spectral range.

In contrast, variation of the HTL thickness has a negligible influence on photogeneration within the device (Fig. 9c). As the P3HT thickness is increased from 50 to 400 nm, the carrier generation rate inside the $Cs_2AgBiBr_6$ absorber remains nearly unchanged, with only a shallow dip observed around 200 nm. The generation rate within the ETL likewise remains essentially constant, while photogeneration inside the HTL itself is negligible across the entire thickness range. These behaviors reflect the fact that photon absorption occurs predominantly within the $Cs_2AgBiBr_6$ absorber, while P3HT, owing to its wider bandgap relative to the absorber, does not contribute appreciably to absorption over the relevant spectral range. The minor reduction in generation at intermediate HTL thickness is attributed to thickness-dependent optical field redistribution within the multilayer stack, which slightly modifies the standing-wave pattern near the absorber/HTL interface rather than introducing additional parasitic absorption.

Overall, these thickness-dependent trends reinforce the earlier wavelength-resolved findings: photogeneration in the CeO$_2$/Cs$_2$AgBiBr$_6$/P3HT architecture is dominated by absorption within the absorber and is primarily governed by optical field distribution and interference effects rather than absorption in the transport layers. Variations in ETL and HTL thickness mainly modulate the spatial redistribution of the optical field and photon flux reaching the absorber, whereas changes in absorber thickness predominantly affect the volumetric averaging of the generation rate rather than the total absorbed photon number. These thickness-dependent generation rates are fed into the Semiconductor module for the subsequent charge transport analysis, allowing a self-consistent evaluation of how optical-geometrical interplay influences device performance − an aspect often simplified or neglected in prior studies.

## 3.4 Electrical Performance

### 3.4.1 Effect of Absorber Thickness and Doping Density

To explore the influence of absorber-layer properties on overall device performance, the thickness and acceptor doping density of the Cs$_2$AgBiBr$_6$ absorber were varied while keeping all other structural and material parameters fixed (Fig. 10a-d). V$_{OC}$ increases with absorber doping density for a given absorber thickness, with a noticeably stronger rate of increase beyond a doping density of approximately $10^{16}$ cm$^{-3}$ (Fig. 10a). When the absorber thickness is varied, V$_{OC}$ increases only weakly at low doping densities ($\leq 10^{16}$ cm$^{-3}$), whereas a more pronounced thickness-dependent enhancement is observed for highly doped absorbers. These trends originate from the evolution of band bending and quasi-Fermi level splitting within the absorber. At low doping densities, weak internal electric fields limit the potential that can be sustained across the absorber, so increasing thickness primarily extends the quasi-neutral region and has little impact on V$_{OC}$. With increasing doping density, stronger band bending at both the CeO$_2$/Cs$_2$AgBiBr$_6$ and Cs$_2$AgBiBr$_6$/P3HT interfaces enhances field-assisted carrier separation and suppresses recombination, allowing quasi-Fermi level splitting to be maintained deeper into the absorber and resulting in a stronger thickness dependence of $V_{OC}$.

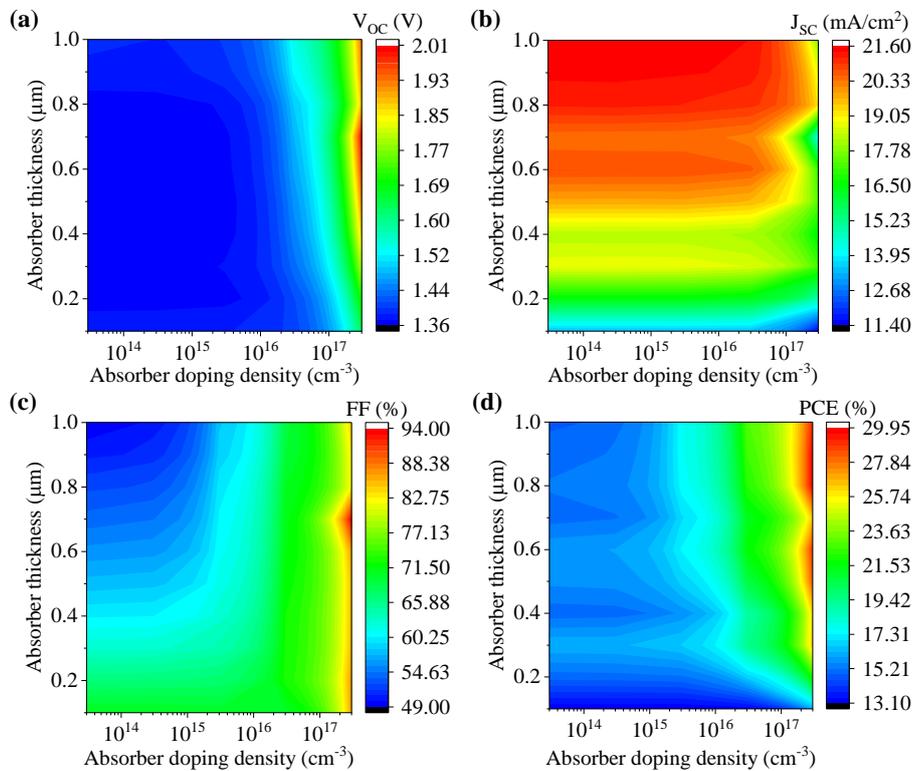

Fig. 10. Effect of thickness and doping density of the absorber, Cs$_2$AgBiBr$_6$ on V$_{OC}$, J$_{SC}$, FF, and PCE of the device. All other material parameters were kept fixed at their default values (Tables 1-3).

$J_{SC}$ exhibits more complex dependencies (Fig. 10b). For a fixed absorber thickness, $J_{SC}$ remains nearly independent of doping density up to approximately $3 \times 10^{16}$ cm$^{-3}$ and decreases at higher doping levels. With increasing absorber thickness, $J_{SC}$ increases at low and moderate doping densities due to enhanced optical absorption, while at the highest doping density studied in this work ($3 \times 10^{17}$ cm$^{-3}$), a non-monotonic thickness dependence is observed, characterized by an initial increase followed by a reduction at intermediate thickness and partial recovery at larger thicknesses. These behaviors reflect the competition between absorption-driven current enhancement and doping-induced recombination losses. At high doping densities, increased free-carrier concentration and defect-assisted recombination shorten carrier lifetimes, reducing current collection efficiency, particularly in thicker absorbers where carriers need to traverse longer transport paths.

For a fixed thickness, FF remains nearly constant with increasing doping density up to approximately $3 \times 10^{14}$ cm$^{-3}$ and increases thereafter (Fig. 10c). When absorber thickness is varied, FF generally decreases with thickness, with the reduction being more pronounced at low doping densities and becoming weaker at higher doping levels. At the highest doping density, FF exhibits a sharp enhancement at an intermediate thickness (~0.7 μm), followed by degradation at larger thicknesses. These trends arise from the interplay between series resistance, recombination, and field-assisted carrier extraction. At low doping densities, limited conductivity leads to increased transport resistance and carrier accumulation in thicker absorbers, reducing FF. Increasing doping density improves absorber conductivity and strengthens the internal electric field, reducing resistive losses and enabling more efficient carrier extraction, which explains the FF enhancement and its partial recovery at high doping and intermediate thickness.

PCE reflects the combined evolution of $V_{OC}$, $J_{SC}$, and FF with absorber doping density and thickness (Fig. 10d). For a given thickness, PCE remains nearly constant at low doping densities and increases at higher doping levels, with the onset of improvement occurring at lower doping densities for thicker absorbers. When absorber thickness is varied, PCE generally increases with thickness, and the rate of enhancement becomes more pronounced at higher doping densities. These trends result from the balance between voltage enhancement, photocurrent generation, and fill-factor evolution. At low doping densities, thickness-induced absorption gains are partially offset by transport and recombination losses, limiting PCE improvement. At higher doping densities, stronger internal electric fields and improved conductivity allow thickness-related absorption gains to translate more effectively into increases in efficiency.

### 3.4.2 Effect of ETL Thickness and Doping Density

Following the absorber-layer analysis, the role of the ETL was examined by varying the thickness and donor doping density of the CeO$_2$ ETL while maintaining all other device parameters constant (Fig. 11a-d). For thick ETLs (> 0.5 μm), $V_{OC}$ increases with ETL doping density for a given thickness, with the enhancement becoming pronounced once the doping density exceeds approximately $10^{18}$ cm$^{-3}$ (Fig. 11a). The increase continues up to ~$2 \times 10^{20}$ cm$^{-3}$, beyond which a reduction in $V_{OC}$ is observed. For thin ETLs (≤ 0.5 μm), $V_{OC}$ remains nearly constant with ETL doping density up to $2 \times 10^{19}$ cm$^{-3}$, followed by a noticeable increase at $2 \times 10^{20}$ cm$^{-3}$ and a slight degradation at higher doping densities. When ETL thickness is varied, $V_{OC}$ increases with thickness up to approximately 0.5 μm for all doping densities. Beyond this thickness, $V_{OC}$ decreases at low ETL doping densities (≤$10^{17}$ cm$^{-3}$), whereas it continues to increase for highly doped ETLs (≥ $10^{18}$ cm$^{-3}$).

These trends originate from the evolution of band bending, electron selectivity, and recombination at the CeO$_2$/Cs$_2$AgBiBr$_6$ interface. At low ETL doping densities, weak band bending and limited electron conductivity reduce the internal electric field and increase interfacial recombination, so increasing ETL thickness beyond an optimal value primarily introduces transport resistance and recombination losses, leading to reduced V$_{OC}$. As ETL doping density increases, stronger band bending enhances electron selectivity and suppresses carrier backflow into the absorber, allowing quasi-Fermi level splitting to be preserved across thicker ETLs and resulting in a stronger thickness-dependent enhancement of V$_{OC}$. At excessively high doping densities, increased free-carrier concentration and defect-assisted recombination within the ETL counteract these benefits, producing the observed reduction in V$_{OC}$.

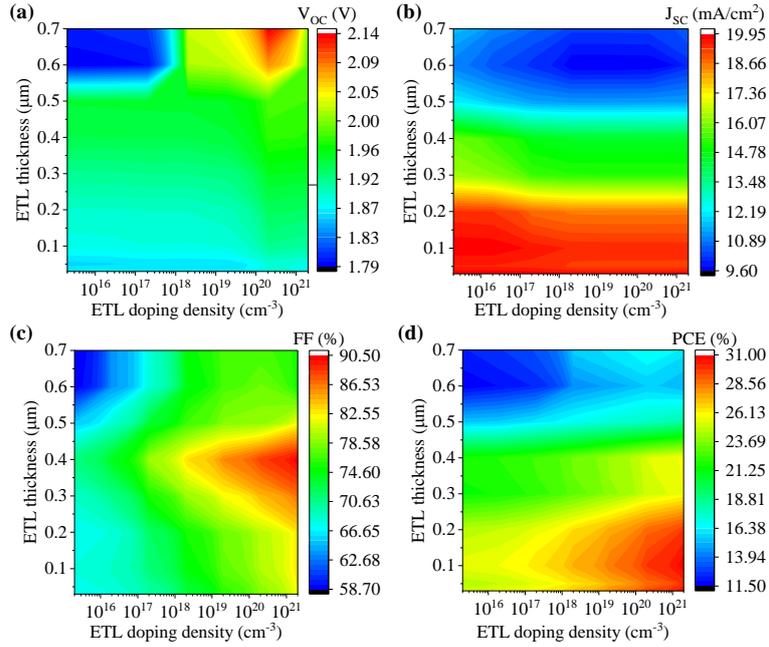

Fig. 11. Effect of thickness and doping density of the ETL (CeO$_2$) on V$_{OC}$, J$_{SC}$, FF, and PCE of the device. The absorber thickness and acceptor doping density were fixed at 800 nm and 3×10$^{17}$ cm$^{-3}$, respectively, which yielded the maximum PCE in section 3.4.1. All other material parameters were kept fixed at their default values (Tables 1-3).

J$_{SC}$ exhibits a comparatively weak dependence on ETL doping density (Fig. 11b). For a fixed ETL thickness, J$_{SC}$ remains nearly independent of ETL doping over most of the investigated range, with only slight reductions observed at high doping densities when the ETL is thick. The maximum variation corresponds to a decrease from approximately 15.84 to 14.5 mA cm$^{-2}$ at an ETL thickness of 0.4 μm. This limited variation indicates that photocurrent generation is primarily governed by optical absorption in the Cs$_2$AgBiBr$_6$ absorber rather than by ETL electronic properties. When ETL thickness is varied, J$_{SC}$ decreases monotonically with increasing thickness, primarily due to enhanced parasitic absorption and increased transport resistance in the CeO$_2$ layer, which limit carrier extraction.

FF increases monotonically with ETL doping density for all ETL thicknesses studied (Fig. 11c). When ETL thickness is varied, FF increases up to an intermediate thickness of approximately 0.4 μm and decreases at larger thicknesses.

These trends arise from the interplay between series resistance, recombination, and field-assisted carrier extraction. Increasing ETL doping density enhances electron conductivity and reduces series resistance, facilitating more efficient carrier extraction and improving FF. Moderate increases in ETL thickness improve junction uniformity and reduce leakage pathways, resulting in initial FF enhancement. Beyond the optimal thickness, however, increased transport resistance and carrier accumulation dominate, leading to FF degradation despite improved ETL conductivity.

For a given ETL thickness, PCE increases monotonically with ETL doping density (Fig. 11d). When ETL thickness is varied, PCE exhibits a non-linear dependence, characterized by an initial decrease at low and moderate thicknesses, followed by a partial recovery at larger thicknesses.

These trends result from the balance between voltage enhancement, photocurrent collection, and fill-factor evolution. At small ETL thicknesses, efficiency gains are obtained due to reduced series resistance and enhanced junction selectivity. As ETL thickness increases, losses in $J_{SC}$ associated with parasitic absorption and transport resistance outweigh gains in $V_{OC}$, leading to PCE degradation. At very large ETL thicknesses and high doping densities, strengthened band bending and improved electron selectivity partially compensate for current losses, resulting in the observed recovery in PCE.

### 3.4.3 Effect of HTL Thickness and Doping Density

Similarly, the influence of HTL properties on device performance was investigated by independently varying the thickness and acceptor doping density of the P3HT HTL while keeping all other parameters fixed (Fig. 12a–d). $V_{OC}$ increases monotonically with HTL doping density for all HTL thicknesses studied, with a noticeably stronger rate of increase at larger HTL thicknesses ($\geq 0.35$ μm) (Fig. 12a). When HTL thickness is varied, $V_{OC}$ exhibits a non-monotonic dependence: it increases initially up to ~0.1 μm, decreases slightly at ~0.15 μm, rises again up to ~0.3 μm, followed by a reduction at ~0.35 μm and a subsequent recovery at larger thicknesses.

These trends originate from the evolution of band bending, hole selectivity, and interfacial recombination at the $Cs_2AgBiBr_6$/P3HT interface. At low HTL doping densities, limited hole conductivity and weak band bending restrict efficient hole extraction, making $V_{OC}$ sensitive to thickness-induced transport losses. Increasing HTL doping strengthens band bending and improves hole selectivity, suppressing interfacial recombination and allowing quasi-Fermi level splitting to be better preserved, particularly in thicker HTLs. The non-monotonic thickness dependence reflects the competing effects of improved junction formation at moderate thicknesses and increased transport resistance and carrier accumulation in thicker P3HT layers. At sufficiently large thicknesses, improved interface quality and stabilized band alignment partially offset transport losses, leading to $V_{OC}$ recovery.

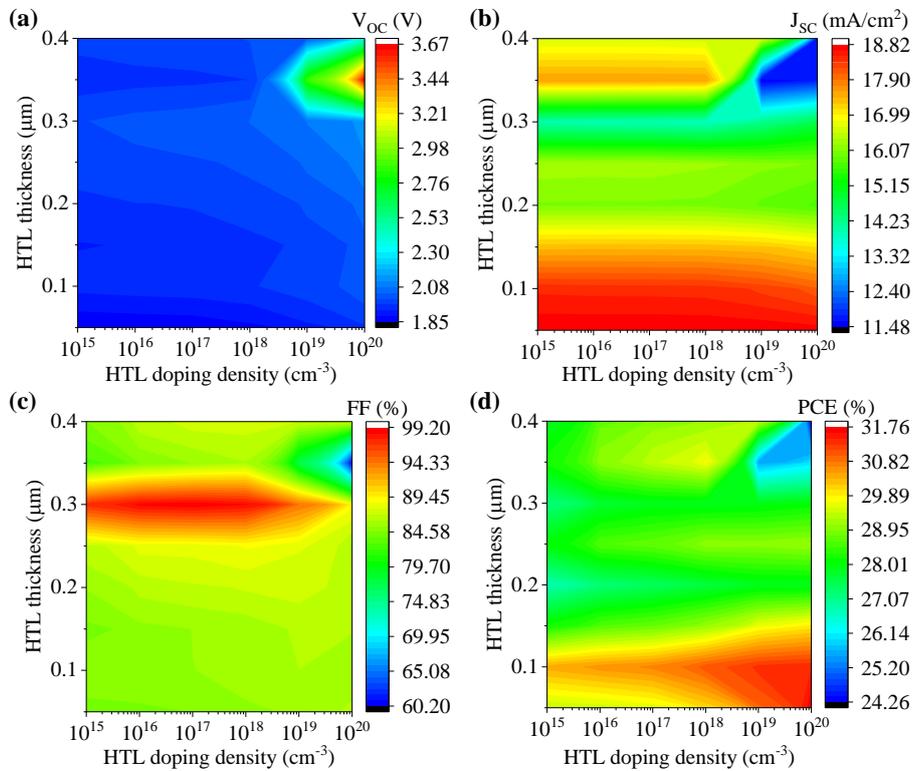

Fig. 12. Effect of thickness and doping density of the HTL (P3HT) on $V_{OC}$, $J_{SC}$, FF, and PCE of the device. The absorber and ETL thicknesses were fixed at 800 nm and 100 nm respectively, while their acceptor and donor doping densities were maintained at $3 \times 10^{17}$ cm$^{-3}$ and $2 \times 10^{21}$ cm$^{-3}$, respectively, which yielded the maximum PCE in section 3.4.2. All other material parameters were kept fixed at their default values (Tables 1-3).

$J_{SC}$ shows a comparatively weak dependence on HTL doping density (Fig. 12b). For HTL thicknesses up to ~0.3 µm, $J_{SC}$ remains nearly independent of doping density, whereas a reduction is observed at higher doping levels ($\geq 10^{19}$ cm$^{-3}$) when the HTL is thick ($\geq 0.35$ µm). When HTL thickness is varied, $J_{SC}$ generally decreases with increasing thickness; however, a noticeable recovery in $J_{SC}$ occurs at larger thicknesses ($\geq 0.3$ µm) when doping density is $\leq 10^{19}$ cm$^{-3}$.

This behavior indicates that photocurrent generation is primarily governed by optical absorption in the Cs$_2$AgBiBr$_6$ absorber, while charge extraction is limited by hole transport in the P3HT layer. The initial reduction in $J_{SC}$ with HTL thickness arises from increased transport resistance and carrier accumulation in thicker layers, which enhance recombination losses. At intermediate doping densities, improved hole conductivity and reduced interfacial recombination partially mitigate these losses at larger thicknesses, leading to the observed recovery in $J_{SC}$.

FF exhibits a moderate dependence on both HTL doping density and thickness (Fig. 12c). For thin HTLs ($\leq 0.25$ µm), FF increases slightly with increasing doping density. At higher thicknesses (0.3–0.35 µm), FF decreases beyond a doping density of ~$10^{18}$ cm$^{-3}$, whereas at the largest thickness studied (0.4 µm), FF shows saturation at high doping densities following an initial increase. When thickness is varied, FF increases up to ~0.3 µm, decreases at ~0.35 µm, and increases again at larger thicknesses.

These trends reflect the balance between series resistance, recombination, and junction uniformity. Moderate increases in HTL thickness improve interfacial coverage and suppress shunt pathways, enhancing FF. However, thicker P3HT layers introduce higher transport resistance and carrier accumulation, leading to FF degradation at intermediate thicknesses. Increasing HTL doping improves hole conductivity and reduces resistive losses, stabilizing FF at large thicknesses and producing saturation behavior at high doping densities.

For HTL thicknesses $\leq 0.3$ µm, PCE increases slightly with doping density, whereas at larger thicknesses ($\geq 0.35$ µm), a reduction in PCE is observed at high doping densities ($\geq 10^{18}$ cm$^{-3}$) (Fig. 12d). When HTL thickness is varied, PCE exhibits a strongly non-monotonic behavior, characterized by alternating regions of enhancement and degradation.

At small HTL thicknesses, efficiency improvements are driven mainly by the improved hole selectivity and reduced series resistance. As thickness increases, JSC losses associated with transport resistance and recombination dominate, reducing PCE. At specific thickness–doping combinations, partial recovery occurs when improved junction quality and stabilized band alignment compensate for transport limitations.

### 3.4.4 Effect of Absorber Defect Density

Prior experimental and theoretical studies have identified a diverse population of defect species in Cs$_2$AgBiBr$_6$, including halide vacancies (V$_{Br}$), cation vacancies (V$_{Ag}$, V$_{Bi}$), and antisite defects (e.g., Ag$_{Bi}$, Bi$_{Ag}$) that introduce trap states within the bandgap and act as non-radiative recombination centers [71], [72]. Therefore, the absorber bulk defect density was treated as a key variable in the present study and varied from $10^{12}$ to $10^{18}$ cm$^{-3}$ to explore the impact on device performance (Fig. 13). All the photovoltaic parameters exhibit a monotonic degradation with increasing defect density. At relatively low defect concentrations ($\leq 10^{14}$ cm$^{-3}$), the parameters decrease only gradually, indicating that non-radiative recombination remains weak and does not yet dominate the overall carrier dynamics. However, once the defect density reaches ~$10^{15}$ cm$^{-3}$ and above, device performance degrades rapidly across all key metrics.

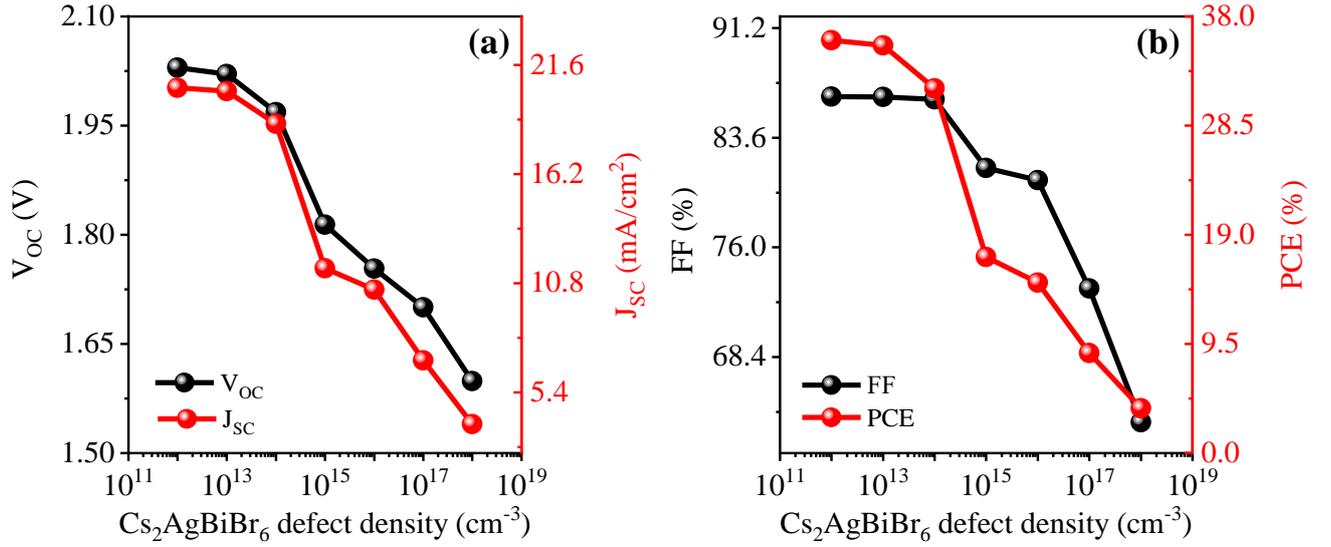

Fig. 13. Effect of bulk defect density in the $Cs_2AgBiBr_6$ on $V_{OC}$, $J_{SC}$, FF, and PCE of the device. The absorber, ETL and HTL thicknesses were fixed at 800 nm, 100 nm and 50 nm, respectively, while their doping densities were maintained at $3 \times 10^{17}$ cm$^{-3}$, $2 \times 10^{21}$ cm$^{-3}$ and $10^{20}$ cm$^{-3}$, respectively, which yielded the maximum PCE in section 3.4.3. All other material parameters were kept fixed at their default values (Tables 1-3).

This transition reflects the onset of defect-assisted SRH recombination as the dominant recombination pathway within the absorber. As the defect density ($N_t$) increases, the carrier lifetimes ($\tau_n$ and $\tau_p$) and diffusion lengths ($L_n$ and $L_p$) are significantly reduced (equation 21-22, [73]), leading to enhanced non-radiative recombination within the absorber.

$$\tau_{n,p} = \frac{1}{\sigma_{n,p} \, N_t \, v_{th}} \tag{21}$$

$$L_{n,p} = \sqrt{\frac{kT}{q} \, \mu_{n,p} \, \tau_{n,p}} \tag{22}$$

where $\sigma$ is the carrier capture cross-section, and $v_{th}$ is the thermal velocity. The resulting reduction in $\tau_{n,p}$ and $L_{n,p}$ directly suppresses the quasi-Fermi level splitting, giving rise to a rapid decline in $V_{OC}$. Simultaneously, intensified bulk recombination reduces the collection probability of photogenerated carriers, causing a measurable decrease in $J_{SC}$. FF is also adversely affected at high defect densities, as increased recombination near the maximum power point degrades diode ideality and flattens the J–V characteristics. As a cumulative outcome of these losses, the PCE deteriorates sharply beyond the critical defect density threshold.

Overall, these results suggest that maintaining defect concentrations below ~$10^{14}$ cm$^{-3}$ is essential to preserve long carrier lifetimes, efficient charge extraction, and high photovoltaic performance in the device.

### 3.4.5 Effect of CTL/Absorber Interface Defect Density

In addition to bulk defects, defects at the $CeO_2/Cs_2AgBiBr_6$ and $P3HT/Cs_2AgBiBr_6$ interfaces can significantly influence device performance. These interface defects primarily originate from interfacial disorder, surface roughness, imperfect band alignment, and incomplete chemical passivation during layer deposition. They act as recombination centers and are commonly described within the SRH formalism through an interface recombination velocity, reflecting their critical influence on carrier extraction and voltage losses.

For an interface defect density $N_{it}$, the SRH interface recombination rate can be expressed as

$$R_{int} = \frac{S_n S_p (np - n_i^2)}{S_n(n + n_1) + S_p(p + p_1)} \tag{23}$$

$$n_1 = n_i \exp\left(\frac{E_t - E_i}{kT}\right), \ p_1 = n_i \exp\left(\frac{E_i - E_t}{kT}\right) \tag{24}$$

where $S_n = \sigma_n v_{th} N_{it}$ and $S_p = \sigma_p v_{th} N_{it}$ are the electron and hole interface recombination velocities, respectively, $\sigma$ denotes the capture cross section, $E_t$ is the energy level of the interface trap, $E_i$ is the intrinsic Fermi level, $n_i$ is the intrinsic carrier concentration, k is the Boltzmann constant and T is the temperature. An increase in $N_{it}$ therefore enhances interface recombination and reduces carrier lifetimes near the interface.

### 3.4.5.1 ETL/Absorber Interface Defect

$V_{OC}$ shows a weakly non-monotonic dependence on $CeO_2/Cs_2AgBiBr_6$ interface defect density (Fig. 14). A slight increase in $V_{OC}$ is observed at very low interface defect densities ($5\times10^7$–$1\times10^8$ cm$^{-2}$) (Fig. 14a). Within the numerical model, this behavior is attributed to minor electrostatic redistribution and changes in local quasi-Fermi level splitting under extremely low interface recombination conditions, rather than to an explicit interface passivation effect. Beyond this narrow regime, at higher defect densities, $V_{OC}$ decreases gradually as enhanced SRH recombination at the interface reduces the quasi-Fermi level splitting.

$J_{SC}$ decreases monotonically with increasing defect density, although the total reduction remains small ($\sim 0.5\%$ up to $5\times10^{12}$ cm$^{-2}$) (Fig. 14a). This weak sensitivity indicates that electron extraction through $CeO_2$ is comparatively resilient to moderate interface recombination. Under short-circuit conditions, carrier collection is dominated by field-assisted transport and favorable conduction-band alignment at the CeO$_2$/Cs$_2$AgBiBr$_6$ interface, which promote rapid carrier sweep-out and limit the residence time of electrons near interfacial recombination centers.

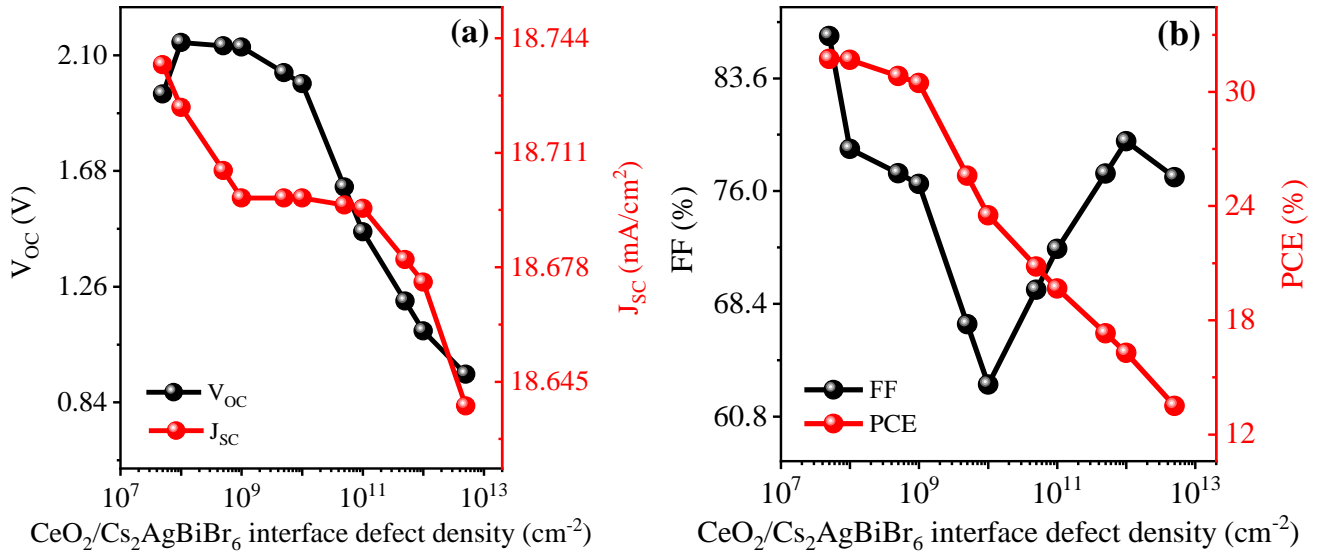

Fig. 14. Effect of defect density at $CeO_2/Cs_2AgBiBr_6$ interface on $V_{OC}$, $J_{SC}$, FF, and PCE of the device. The absorber, ETL and HTL thicknesses were fixed at 800 nm, 100 nm and 50 nm, respectively, while their doping densities were maintained at $3\times10^{17}$ cm$^{-3}$, $2\times10^{21}$ cm$^{-3}$ and $10^{20}$ cm$^{-3}$, respectively, which yielded the maximum PCE in section 3.4.3. All other material parameters were kept fixed at their default values (Tables 1-3).

FF displays a more complex response (Fig. 14b). FF initially decreases with increasing interface defect density up to approximately $10^{10}$ cm$^{-2}$, reflecting enhanced recombination losses near the maximum power point. Interestingly, a partial recovery in FF is observed at intermediate defect densities, before a renewed decline occurs once the defect density exceeds $10^{12}$ cm$^{-2}$. This non-monotonic trend suggests a competition between recombination-induced losses and changes in interfacial charge transport, where moderate defect densities may transiently modify local electric fields or carrier selectivity, while high defect densities ultimately dominate through recombination-driven degradation.

As a cumulative consequence of these effects, the PCE decreases continuously with increasing ETL/absorber interface defect density (Fig. 14b), confirming that interface recombination at the electron-extraction side progressively undermines device performance.

### 3.4.5.2 HTL/Absorber Interface Defect

In contrast, defects at the P3HT/Cs$_2$AgBiBr$_6$ interface exert a stronger influence on device performance (Fig. 15). V$_{OC}$ decreases steadily with increasing defect density and saturates beyond $\sim 10^{11}$ cm$^{-2}$ (Fig. 15a), consistent with recombination-limited quasi-Fermi level splitting once interface SRH recombination dominates.

J$_{SC}$ also decreases with increasing defect density, with a pronounced drop emerging once the interface defect density reaches $5 \times 10^{12}$ cm$^{-2}$ (Fig. 15a). The reduction in J$_{SC}$ is attributed to enhanced trap-assisted recombination at the P3HT/Cs$_2$AgBiBr$_6$ interface, where photogenerated holes accumulate under operation, increasing their residence time near interfacial defect states and suppressing efficient carrier extraction.

FF follows a non-monotonic trend, exhibiting an initial decline at low defect densities, a weak recovery at intermediate densities, and a sharp degradation beyond $\sim 10^{12}$ cm$^{-2}$ (Fig. 15b). At high defect concentrations, enhanced interface recombination significantly distorts the diode ideality and increases recombination losses near the maximum power point.

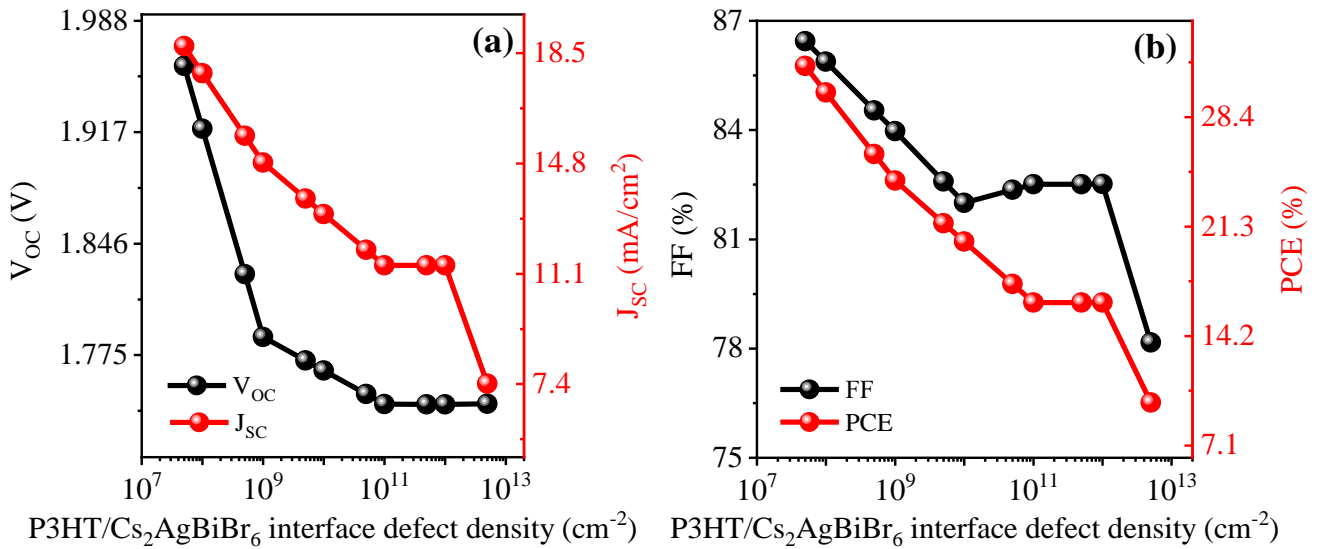

Fig. 15. Effect of defect density at P3HT/Cs$_2$AgBiBr$_6$ interface on V$_{OC}$, J$_{SC}$, FF, and PCE of the device. The absorber, ETL and HTL thicknesses were fixed at 800 nm, 100 nm and 50 nm, respectively, while their doping densities were maintained at $3 \times 10^{17}$ cm$^{-3}$, $2 \times 10^{21}$ cm$^{-3}$ and $10^{20}$ cm$^{-3}$, respectively, which yielded the maximum PCE in section 3.4.3. All other material parameters were kept fixed at their default values (Tables 1-3).

Consequently, the PCE decreases with increasing HTL/absorber interface defect density (Fig. 15b), remaining nearly constant only over a narrow intermediate regime before undergoing a sharp decline at high defect densities.

The results of this section highlight that effective control of interfacial defect densities is a key requirement for maintaining efficient carrier extraction and achieving high-performance Cs$_2$AgBiBr$_6$-based solar cells.

### 3.4.6 Effect of Direct and Auger Recombination Coefficient

When the radiative (direct) and Auger recombination coefficients of the Cs$_2$AgBiBr$_6$ absorber are varied simultaneously, a clear threshold-dependent degradation of device performance is observed (Fig. 16). For low recombination coefficients, all photovoltaic parameters remain nearly unchanged, indicating that recombination losses are not rate-limiting in this regime. However, once either recombination coefficient increases beyond a critical level, a pronounced deterioration in device performance occurs.

At low radiative and Auger recombination coefficients, photogenerated carriers in the Cs$_2$AgBiBr$_6$ absorber possess sufficiently long lifetimes to be efficiently separated and extracted by the built-in electric fields at the CeO$_2$/Cs$_2$AgBiBr$_6$ and Cs$_2$AgBiBr$_6$/P3HT interfaces. Under these conditions, quasi-Fermi level splitting is well preserved, leading to high V$_{OC}$, while efficient carrier collection sustains both J$_{SC}$ and FF.

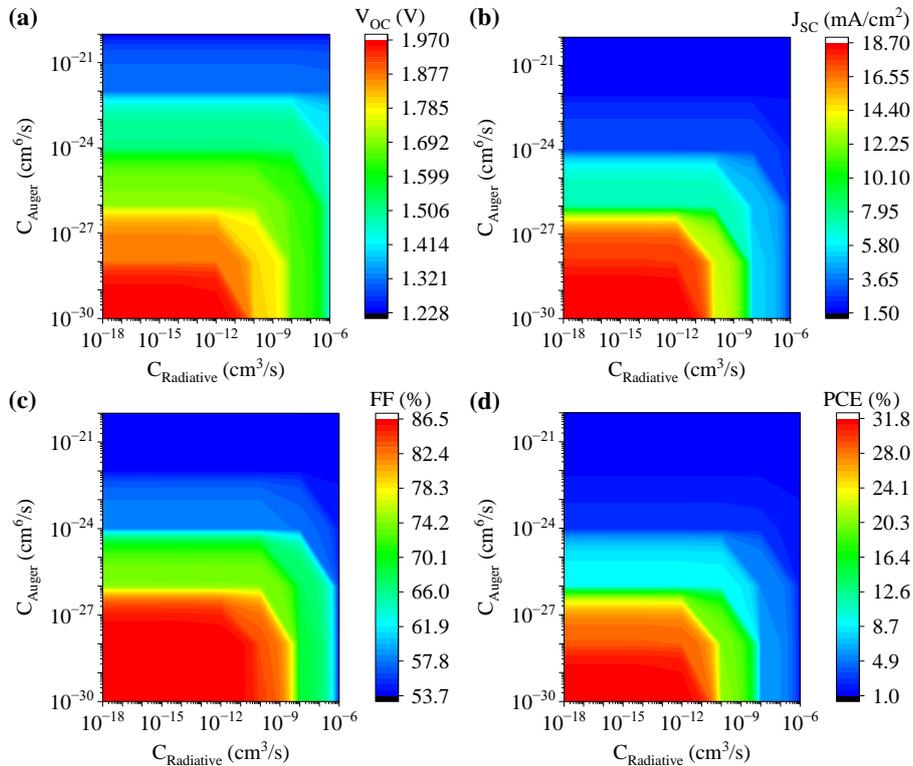

Fig. 16. Effect of radiative and Auger recombination coefficients on V$_{OC}$, J$_{SC}$, FF, and PCE of the device. All other material parameters were kept fixed at their default values (Tables 1-3), and the Auger coefficients were assumed identical for electrons and holes. The absorber, ETL and HTL thicknesses were fixed at 800 nm, 100 nm and 50 nm, respectively, while their doping densities were maintained at 3×10$^{17}$ cm$^{-3}$, 2×10$^{21}$ cm$^{-3}$ and 10$^{20}$ cm$^{-3}$, respectively, which yielded the maximum PCE in section 3.4.3. All other material parameters were kept fixed at their default values (Tables 1-3).

As the radiative recombination coefficient increases beyond its threshold, band-to-band recombination losses become significant. Although radiative recombination is an intrinsic process, an excessively high radiative coefficient reduces the effective carrier lifetime and limits the achievable quasi-Fermi level splitting, resulting in a noticeable reduction in $V_{OC}$ (Fig. 16a). The accompanying increase in bulk recombination also lowers carrier collection efficiency, contributing to simultaneous declines in $J_{SC}$ and FF (Fig. 16b,c).

Auger recombination has an even more detrimental impact at higher carrier densities due to its strong dependence on carrier concentration. Once the Auger recombination coefficient exceeds $\sim 10^{-29}\,\mathrm{cm^6\,s^{-1}}$, three-carrier recombination processes dominate, particularly in regions of strong band bending near the junction where carrier densities are highest. This leads to rapid carrier annihilation, severely reducing carrier lifetime and diffusion length. As a consequence, the performance parameters degrade markedly.

The cumulative effect of these recombination losses is clearly reflected in the PCE (Fig. 16d). PCE exhibits a pronounced decline once either recombination coefficient exceeds its critical threshold. Based on the observed trends, optimal device performance is achieved when the radiative recombination coefficient remains below $\sim 10^{-11}\,\mathrm{cm^3\,s^{-1}}$ and the Auger recombination coefficient remains below $\sim 10^{-29}\,\mathrm{cm^6\,s^{-1}}$.

### 3.4.7 Effect of Temperature

The temperature dependence of the photovoltaic performance of the device was investigated over a wide temperature range from 200 to 500 K to assess the interplay between thermal activation, charge transport, and recombination (Fig. 17). All device metrics exhibit non-monotonic temperature dependence, indicating a transition between transport-limited and recombination-limited operating regimes.

$V_{OC}$ initially increases with temperature up to approximately 300 K, followed by a gradual decrease up to ~350 K and saturation at higher temperatures. This behavior is consistent with the temperature dependence of the quasi-Fermi level splitting (equation 16). At low temperatures (200–300 K), enhanced thermal activation of the carriers and reduced interfacial barriers lead to a positive temperature coefficient, $\left(\frac{dV_{OC}}{dT}\right)_{T<300K} > 0$. At higher temperatures, the exponential increase of $J_0$ dominates [74],

$$J_0 \propto T^3 \exp\left(-\frac{E_g}{kT}\right) \tag{25}$$

resulting in a negative temperature coefficient, $\left(\frac{dV_{OC}}{dT}\right)_{T>300K} < 0$, as recombination and effective bandgap narrowing reduce the attainable voltage. The eventual saturation of $V_{OC}$ above ~350 K indicates a recombination-controlled regime in which further thermal activation no longer alters the quasi-Fermi level splitting.

$J_{SC}$ shows a similarly non-monotonic temperature dependence. $J_{SC}$ increases slightly from 200 K to ~275 K due to enhanced carrier mobility and improved charge collection efficiency. With further temperature increase up to ~325 K, $J_{SC}$ decreases as carrier–phonon scattering and thermally activated recombination increasingly limit carrier transport. At higher temperatures, $J_{SC}$ becomes nearly temperature independent, suggesting that optical generation and transport losses reach a dynamic balance that is weakly sensitive to further thermal variation.

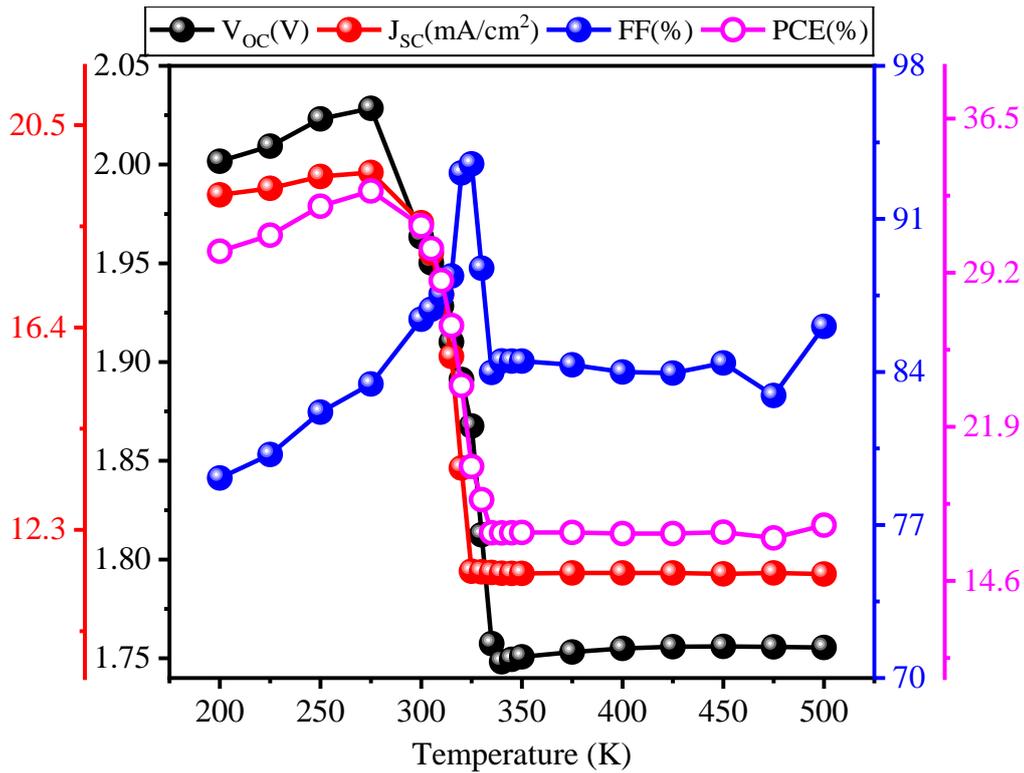

Fig. 17. Effect of temperature (T) on $V_{OC}$, $J_{SC}$, FF, and PCE of the device. The absorber, ETL and HTL thicknesses were fixed at 800 nm, 100 nm and 50 nm, respectively, while their doping densities were maintained at $3\times10^{17}$ cm$^{-3}$, $2\times10^{21}$ cm$^{-3}$ and $10^{20}$ cm$^{-3}$, respectively, which yielded the maximum PCE in section 3.4.3. All other material parameters were kept fixed at their default values (Tables 1-3).

FF shows an initial improvement with temperature, reaching a maximum near ~325 K, followed by a gradual decline at higher temperatures and eventual saturation above ~350 K. The initial increase in FF can be attributed to reduced series resistance and improved carrier transport with increasing temperature. At higher temperatures, however, increased recombination and enhanced resistive losses reduce the squareness of the current–voltage characteristics, leading to FF degradation.

As the cumulative metric, PCE closely tracks the combined evolution of $V_{OC}$, $J_{SC}$, and FF. PCE increases initially with temperature, reaching a maximum near ~275 K, driven by the simultaneous enhancement of $V_{OC}$, a modest increase in $J_{SC}$, and improved FF. At higher temperatures, the concurrent reduction in $V_{OC}$ and $J_{SC}$ dominates, resulting in a progressive decrease in PCE up to ~340 K, followed by a plateau at higher temperatures. This saturation reflects the establishment of a thermally limited operating regime in which further increases in temperature do not substantially alter the balance between photogeneration, transport, and recombination.

Overall, the temperature-dependent behavior underscores the critical role of thermally activated recombination in limiting device performance at elevated temperatures, while also highlighting an optimal operating window near room temperature where transport, recombination, and interfacial energetics are most favorably balanced in the device.

**3.4.8 Effect of Incident Light Intensity**

The operational response of the device under varying illumination conditions was evaluated by examining the dependence of the photovoltaic parameters on incident light intensity (Fig. 18), an essential consideration given the substantial fluctuations in irradiance between outdoor and indoor environments and under different weather conditions. $V_{OC}$ exhibits a logarithmic increase with increasing light intensity, in agreement with the theoretical relationship described by equation (16). Since $J_{SC}$ scales linearly with illumination, $V_{OC}$ increases logarithmically. Physically, higher photon flux generates a larger population of photocarriers, leading to enhanced quasi-Fermi level splitting within the absorber and a corresponding rise in $V_{OC}$.

$J_{SC}$ increases linearly with incident light intensity ($\frac{\partial (Jsc)}{\partial (Intensity)}$ = constant = 187.52 mA/W), reflecting the direct proportionality between photon flux and the rate of photogenerated electron–hole pairs [75]. This proportionality indicates that photocarrier generation and collection remain efficient over the investigated intensity range, with no evidence of current saturation or transport-induced limitations.

A slight yet systematic increase in the FF is observed with increasing illumination intensity. This trend can be attributed to improved diode rectification and a reduced relative influence of resistive losses at higher photogenerated current densities, which together enhance the curvature of the J–V characteristics near the maximum power point.

Consequently, the PCE increases monotonically with illumination intensity, driven primarily by the linear enhancement of $J_{SC}$ and the logarithmic rise in $V_{OC}$, with secondary contributions from FF improvement. Notably, the PCE remains substantial under low-intensity illumination and exhibits stable behavior at higher irradiances, indicating the absence of recombination-induced rollover or resistive degradation across the investigated range.

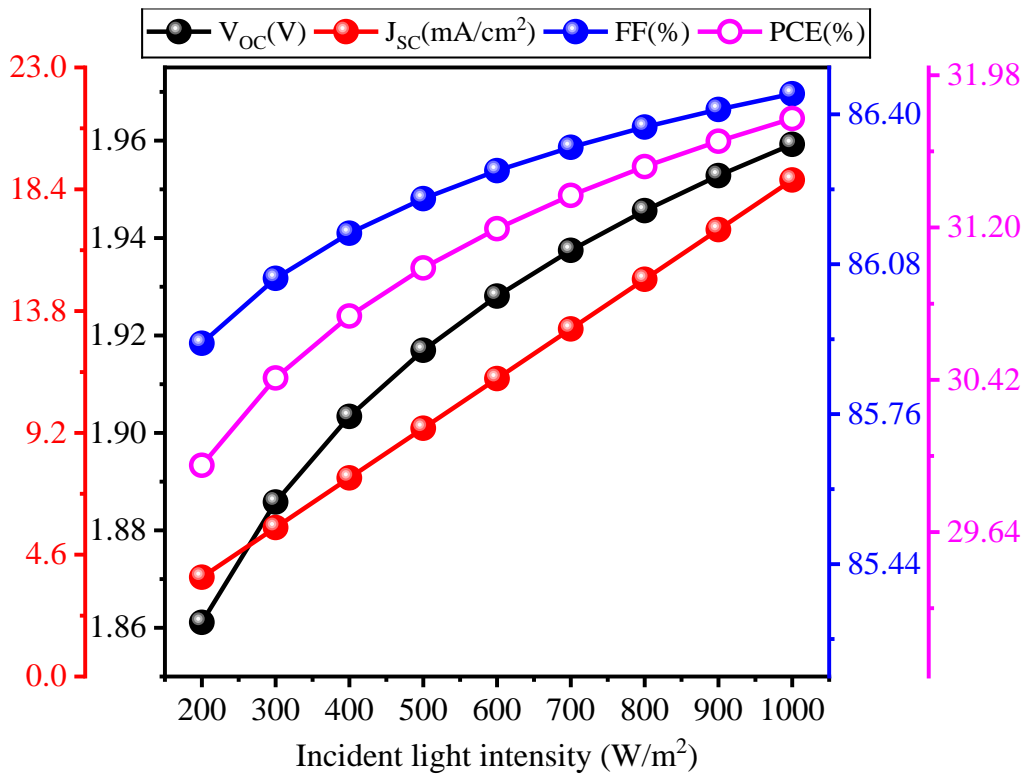

Fig. 18. Effect of incident light intensity on $V_{OC}$, $J_{SC}$, FF, and PCE of the device. The absorber, ETL and HTL thicknesses were fixed at 800 nm, 100 nm and 50 nm, respectively, while their doping densities were maintained at $3\times10^{17}$ cm$^{-3}$, $2\times10^{21}$ cm$^{-3}$ and $10^{20}$ cm$^{-3}$, respectively, which yielded the maximum PCE in section 3.4.3. All other material parameters were kept fixed at their default values (Tables 1-3).

## 3.5 Optimized Device

Based on the preceding optical and electrical analyses, the device parameters were systematically tuned to identify an optimized device configuration under standard operating conditions (AM1.5G illumination at 1000 W m$^{-2}$ and 300 K). The optimized FTO/CeO$_2$/Cs$_2$AgBiBr$_6$/P3HT/Au device delivers a V$_{OC}$ of 1.96 V, J$_{SC}$ of 18.75 mA cm$^{-2}$, FF of 86.44%, and PCE of 31.76%. The corresponding J–V and P–V characteristics and external quantum efficiency (EQE) response are presented in Fig. 19.

To place these results in context, the optimized device performance was compared with recently reported Cs$_2$AgBiBr$_6$-based solar cells from both experimental and numerical studies published between 2022 and 2026 (Table 6). Experimentally reported efficiencies span a range from 0.19% to 6.37%, while simulation studies predicted efficiencies between 5.56% and 27.78%. The simulated optimal efficiency obtained in the present work exceeds the values reported in these studies, underscoring the potential of the proposed device architecture and the comprehensive 3D physics-based optimization strategy. It should be emphasized that the reported value represents the best-case simulated performance under idealized conditions.

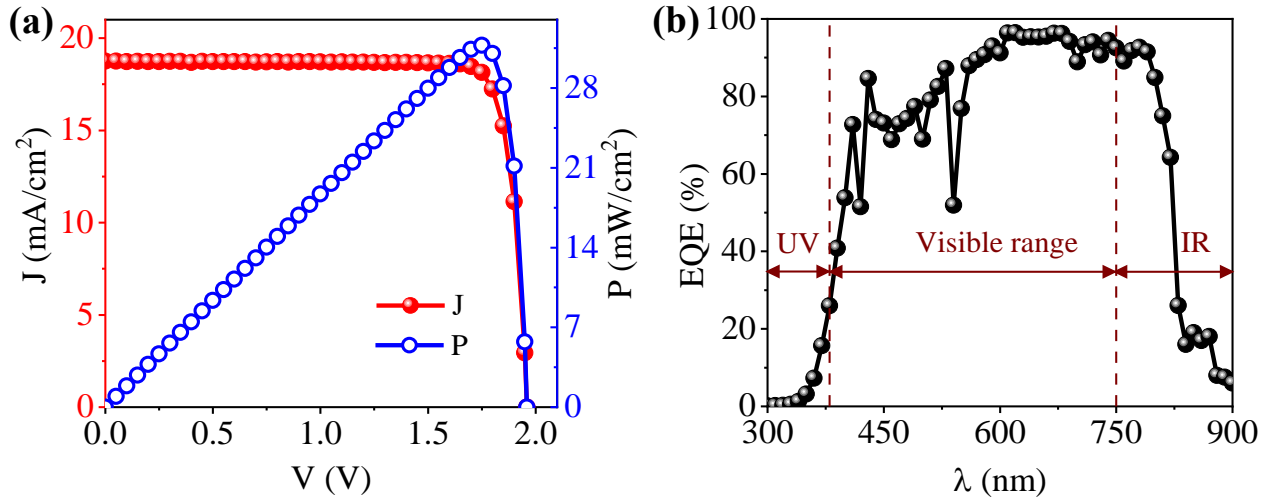

Fig. 19. (a) J–V and P (power)–V characteristics, and (b) EQE of the optimized FTO/CeO$_2$/Cs$_2$AgBiBr$_6$/P3HT/Au device. EQE was calculated using the formula [76]: $EQE(\lambda) = \frac{J(\lambda)}{q\Phi_{ph}(\lambda)}$, where J is the electron flux, in the form of electrical current (at short circuit) and $\Phi_{ph}$ is the incident photon flux. In panel (b), the AM1.5G spectrum is shown in the background in light brown, with values corresponding to normalized spectral irradiance. The UV (ultra-violet) region (300–380 nm), visible range (380–750 nm), and IR (infra-red) region (>750 nm) are indicated for reference. Thickness of different layers were set as: CeO$_2$ = 100 nm, Cs$_2$AgBiBr$_6$ = 800 nm, P3HT = 50 nm; doping densities were set as: CeO$_2$ = 2×10$^{21}$ cm$^{-3}$, Cs$_2$AgBiBr$_6$ = 3×10$^{17}$ cm$^{-3}$, P3HT = 10$^{20}$ cm$^{-3}$. All other material parameters were kept fixed at their default values, as listed in (Tables 1-3).

Table 6. Comparison of photovoltaic performance of the proposed Cs$_2$AgBiBr$_6$ absorber–based solar cell with recent experimental and simulation studies (2022–2026).

| Device structure | V$_{OC}$ (V) | J$_{SC}$ (mA/cm$^2$) | FF (%) | PCE (%) | Reference | Year |
|---|---|---|---|---|---|---|
| **Experimental works** | | | | | | |
| FTO/SnO$_2$/Cs$_2$AgBiBr$_6$/Carbon | 0.69 | 1.71 | 47.9 | 0.54 | [77] | 2023 |
| FTO/TiO$_2$/Cs$_2$AgBiBr$_6$/Carbon | 1.22 | 3.04 | 62.0 | 2.03 | [78] | 2023 |
| ITO/SnO$_2$/Cs$_2$AgBiBr$_6$/Spiro-OMeTAD/Au | 0.89 | 11.4 | 55.6 | 6.37 | [28] | 2022 |
| SnO$_2$/Cs$_2$AgBiBr$_6$/Spiro-OMeTAD | 0.92 | 11.4 | 60.93 | 6.37 | [28] | 2022 |
| PTB7/Au/FTO/c-TiO$_2$/m-TiO$_2$/Cs$_2$AgBiBr$_6$/Spiro-OMeTAD/Ag | 1.03 | 5.14 | 55.4 | 3.07 | [79] | 2022 |
| FTO/c-TiO$_2$/Cs$_2$AgBiBr$_6$/Carbon | 1.29 | 0.49 | 30.0 | 0.19 | [80] | 2022 |
| FTO/c-TiO$_2$/m-TiO$_2$/Cs$_2$AgBiBr$_6$/(PEA)$_4$AgBiBr$_8$/Spiro-OMeTAD/Au | 1.07 | 3.5 | 66.0 | 2.47 | [81] | 2022 |
| TiO$_2$/Cs$_2$AgBiBr$_6$/Spiro-OMeTAD | 1.08 | 3.55 | 64.0 | 2.46 | [81] | 2022 |
| FTO/c-TiO$_2$/m-TiO$_2$/Cs$_2$AgBiBr$_6$/Carbon | 1.2 | 2.61 | 71.0 | 2.22 | [82] | 2022 |
| FTO/c-TiO$_2$/m-TiO$_2$/Cs$_2$AgBiBr$_6$/MoS$_2$/carbon | 1.1 | 6.12 | 72.0 | 4.17 | [83] | 2022 |
| FTO/c-TiO$_2$/m-TiO$_2$/Cs$_2$AgBiBr$_6$/poly[[$_{4,8}$-bis[$_5$-($_2$-ethylhexyl)-$_2$-thienyl]benzo[$_{1,2}$-b':$_{4,5}$-b'] | 1.28 | 3.34 | 77.5 | 3.31 | [84] | 2022 |
| FTO/c-TiO$_2$/m-TiO$_2$/Cs$_2$AgBiBr$_6$/Spiro-OMeTAD/Ag | 1.05 | 1.77 | 71.7 | 1.33 | [85] | 2022 |
| FTO/c-TiO$_2$/m-TiO$_2$/Cs$_2$AgBiBr$_6$/Spiro-OMeTAD/Ag | 1.23 | 2.38 | 71.0 | 2.09 | [86] | 2022 |
| FTO/c-TiO$_2$/m-TiO$_2$/D$_{419}$-Cs$_2$AgBiBr$_6$@Ti$_3$C$_2$Tx/Spiro-OMeTAD/Ag | 0.72 | 8.85 | 70.1 | 4.47 | [87] | 2022 |
| FTO/m-TiO$_2$/Cs$_2$AgBiBr$_6$/SnS QDs/carbon | 1.02 | 3.74 | 51.0 | 1.95 | [88] | 2022 |
| FTO/TiO$_2$/Cs$_2$Ag$_1$-xZnxBiBr$_6$/Carbon | 1.0 | 4.23 | 51.0 | 2.16 | [89] | 2022 |
| ITO/SnO$_2$/Cs$_2$AgBiBr$_6$/P$_3$HT/MoO$_3$/PTAA/Au | 1.06 | 2.8 | 65.0 | 1.94 | [90] | 2022 |
| **Simulation works** | | | | | | |
| FTO/TiO$_2$/Cs$_2$AgBiBr$_6$/Cu$_2$O/Au | 1.71 | 23.33 | 76.24 | 27.78 | [38] | 2026 |
| FTO/SnO$_2$/Cs$_2$AgBiBr$_6$/P$_3$HT/Al | 1.79 | 13.26 | 75.41 | 17.56 | [91] | 2025 |
| FTO/TiO$_2$/Cs$_2$AgBiBr$_6$//CuI/Au | 1.04 | 8.54 | 62.37 | 5.56 | [92] | 2025 |
| FTO/TiO$_2$/Cs$_2$AgBiBr$_6$/Cu$_2$O/Au | 1.33 | 8.78 | 64.96 | 7.6 | [92] | 2025 |
| FTO/TiO$_2$/Cs$_2$AgBiBr$_6$/CuSCN/Au | 1.14 | 8.6 | 63.71 | 6.28 | [92] | 2025 |
| FTO/TiO$_2$/Cs$_2$AgBiBr$_6$/GQDs/Au | 0.814 | 24.55 | 76.69 | 15.33 | [93] | 2025 |
| FTO/TiO$_2$/Cs$_2$AgBiBr$_6$/Spiro-OMEOTED/Au | 1.06 | 8.55 | 62.53 | 5.68 | [92] | 2025 |
| FTO/ZnO/Cs$_2$AgBiBr$_6$/CBTS/Au | 1.42 | 19.978 | 89.958 | 25.562 | [36] | 2025 |
| FTO/ZnO/Cs$_2$AgBiBr$_6$/CFTS/Au | 1.13 | 22.08 | 62.93 | 15.65 | [36] | 2025 |
| ITO/TiO$_2$/Cs$_2$AgBiBr$_6$/Cu$_2$O/Au | 1.11 | 16.54 | 82.43 | 15.11 | [94] | 2025 |
| FTO/WS$_2$/Cs$_2$AgBiBr$_6$/Spiro-OMeTAD/Au | 1.5 | 18.86 | 85.02 | 24.16 | [95] | 2025 |

| Device structure | $V_{OC}$ (V) | $J_{SC}$ (mA/cm$^2$) | FF (%) | PCE (%) | Reference | Year |
|---|---|---|---|---|---|---|
| FTO/TiO$_2$/IDL/Cs$_2$AgBiBr$_6$ (gradient)/C | 1.69 | 12.11 | 88.79 | 18.21 | [96] | 2025 |
| FTO/AZnO/Cs$_2$AgBiBr$_6$/CNTS/Au | 1.375 | 21.38 | 79.93 | 23.5 | [48] | 2024 |
| FTO/CdZnS/Cs$_2$AgBiBr$_6$/CNTS/Au | 1.372 | 21.41 | 78.82 | 23.15 | [48] | 2024 |
| FTO/CdZnS/Cs$_2$AgBiBr$_6$/CNTS/Au | 1.377 | 21.42 | 79.09 | 23.34 | [97] | 2024 |
| FTO/LBSO/Cs$_2$AgBiBr$_6$/CNTS/Au | 1.373 | 21.4 | 79.72 | 23.42 | [48] | 2024 |
| FTO/Nb$_2$O$_5$/Cs$_2$AgBiBr$_6$/CNTS/Au | 1.376 | 21.41 | 78.87 | 23.23 | [48] | 2024 |
| FTO/SnO$_2$/Cs$_2$AgBiBr$_6$/CuI/Au | 1.04 | 11.79 | 66.41 | 8.15 | [98] | 2024 |
| FTO/ZnO/Cs$_2$AgBiBr$_6$/CNTS/Au | 1.37 | 21.38 | 79.93 | 23.5 | [48] | 2024 |
| FTO/AZO/Cs$_2$AgBiBr$_6$/ZnTe | 1.5 | 21.35 | 83.87 | 26.89 | [35] | 2024 |
| FTO/SnO$_2$/Cs$_2$AgBiBr$_6$/Spiro-OMeTAD/Cu | 1.787 | 8.648 | 36.38 | 5.62 | [99] | 2024 |
| FTO/WO$_3$/Cs$_2$AgBiBr$_6$/NiO/Cu-doped C | 1.4 | 20.0 | 74.76 | 20.93 | [100] | 2024 |
| FTO/C$_{60}$/Cs$_2$AgBiBr$_6$/MoS$_2$/Pt | 0.84 | 32.28 | 85.77 | 23.49 | [101] | 2023 |
| FTO/PCBM/Cs$_2$AgBiBr$_6$/NiOx/Au | 2.49 | 8.17 | 76.16 | 15.49 | [1] | 2023 |
| FTO/PCBM/Cs$_2$AgBiBr$_6$/NiOx/Au | 2.49 | 11.49 | 77.03 | 21.92 | [1] | 2023 |
| FTO/SnO$_2$/Cs$_2$AgBiBr$_6$/Spiro-OMeTAD/Au | 1.3 | 17.44 | 62.59 | 14.29 | [102] | 2023 |
| FTO/ZnO/Cs$_2$AgBiBr$_6$/Cu$_2$O/Au | 0.86 | 16.56 | 53.84 | 7.66 | [103] | 2023 |
| FTO/ZnO/Cs$_2$AgBiBr$_6$/NiO/Au | 1.29 | 20.69 | 81.72 | 21.88 | [103] | 2023 |
| ITO/CdS/Cs$_2$AgBiBr$_6$/CuAlO$_2$/Pt | 1.642 | 4.913 | 88.74 | 7.16 | [104] | 2023 |
| ITO/Spiro-OMeTAD/Cs$_2$AgBiBr$_6$/SnO$_2$/Au | 1.278 | 23.3 | 88.21 | 26.3 | [105] | 2023 |
| SnO$_2$/Cs$_2$AgBiBr$_6$/Me$_4$PAC | 1.527 | 10.42 | 75.66 | 12.08 | [106] | 2023 |
| Zn$_{0.75}$Mg$_{0.25}$O/Cs$_2$AgBiBr$_6$/Cu-doped NiO$_x$ | 1.68 | 14.81 | 80.54 | 20.06 | [107] | 2023 |
| Zn$_{0.75}$Mg$_{0.25}$O/Cs$_2$AgBiBr$_6$/Cu$_2$O | 1.39 | 14.8 | 77.47 | 16.07 | [107] | 2023 |
| Zn$_{0.75}$Mg$_{0.25}$O/Cs$_2$AgBiBr$_6$/MASnBr$_3$ | 1.66 | 14.81 | 80.42 | 19.86 | [107] | 2023 |
| Zn$_{0.75}$Mg$_{0.25}$O/Cs$_2$AgBiBr$_6$/NiO$_x$ | 1.45 | 14.8 | 78.43 | 16.84 | [107] | 2023 |
| Zn$_{0.75}$Mg$_{0.25}$O/Cs$_2$AgBiBr$_6$/P3HT | 1.24 | 14.79 | 73.64 | 13.51 | [107] | 2023 |
| Zn$_{0.75}$Mg$_{0.25}$O/Cs$_2$AgBiBr$_6$/Spiro-OMeTAD | 1.44 | 14.8 | 78.0 | 16.72 | [107] | 2023 |
| FTO/TiO$_2$/Cs$_2$AgBiBr$_6$/Spiro-OMeTAD/Au | 0.9 | 11.09 | 66.61 | 6.68 | [108] | 2022 |
| Ni/NiO$_x$/Cs$_2$AgBiBr$_6$/TiO$_2$/FTO/Al | 1.33 | 21.46 | 88.53 | 25.38 | [109] | 2022 |
| ZnO/Cs$_2$AgBiBr$_6$/NiO | 1.29 | 20.69 | 81.72 | 21.88 | [103] | 2022 |
| FTO/CeO$_2$/ Cs$_2$AgBiBr$_6$/P3HT/Au | 1.96 | 18.75 | 86.44 | 31.76 | **This work** | 2026 |

## 4. Conclusion

In this work, a physics-based three-dimensional numerical framework, grounded in Maxwell's electrodynamics and semiconductor drift-diffusion theory, was developed to investigate the performance-limiting factors and optimization pathways for $Cs_2AgBiBr_6$-based lead-free double perovskite solar cells. A systematic screening of 25 ETL–HTL combinations, constructed from five representative electron transport layers ($CeO_2$, AZO, $C_{60}$, PCBM, and $WS_2$) and five hole transport layers (P3HT, CuO, NiO, PEDOT:PSS, and $MoO_3$), identified the $CeO_2$/P3HT pairing as the most promising architecture, delivering the highest power conversion efficiency among all evaluated configurations. Building on the FTO/$CeO_2$/$Cs_2AgBiBr_6$/P3HT/Au stack, comprehensive optical and electrical analyses were conducted, covering optimization of layer thicknesses and doping densities, together with quantitative assessment of bulk and interfacial defect states, to resolve the coupled photogeneration, carrier transport, and recombination dynamics governing device operation. Optical analyses revealed strongly front-loaded photogeneration, with peak absorption localized near the ETL/absorber interface and generation rates on the order of ~$10^{20}$ $cm^{-3}$ $s^{-1}$ across the 400–800 nm spectral range. Importantly, generation profiles were calculated for each absorber, ETL, and HTL thickness variation, enabling realistic capture of optical redistribution and interference effects that are commonly neglected in typical device modeling. Systematic variation of absorber thickness and doping density demonstrated that $V_{OC}$ and FF dominates efficiency gains at higher doping levels ($\geq 3 \times 10^{15}$ $cm^{-3}$), whereas photocurrent improvements associated with increased thickness are progressively offset by transport and recombination losses at excessive doping or thickness. Engineering of the $CeO_2$ layer revealed an optimal doping window (> $10^{19}$ $cm^{-3}$) that enhances electron selectivity and device efficiency, while excessive ETL thickness (> 0.5 μm) introduces parasitic absorption and transport resistance, reducing $J_{SC}$ and PCE. Optimization of the P3HT layer exposed a pronounced non-monotonic dependencies of $V_{OC}$, FF, and PCE on both thickness and acceptor density, with higher efficiencies achieved at higher doping levels (> $10^{19}$ $cm^{-3}$) and reduced thicknesses (< 150 nm). Defect physics emerged as a dominant determinant of achievable performance. Bulk defect densities in $Cs_2AgBiBr_6$ exceeding ~$10^{14}$ $cm^{-3}$ drive a transition toward recombination-limited operation, sharply degrading photovoltaic metrics through reduced carrier lifetime and diffusion length. Interface defects exhibit asymmetric impact: the $CeO_2$/$Cs_2AgBiBr_6$ interface shows relative tolerance to moderate trap densities ($\leq 10^9$ $cm^{-2}$), whereas the P3HT/$Cs_2AgBiBr_6$ interface demonstrates pronounced sensitivity, with continuous PCE degradation across the investigated defect range. Analysis of intrinsic recombination pathways revealed distinct threshold behavior. Device performance remains stable for radiative recombination coefficients below ~$10^{-11}$ $cm^3$ $s^{-1}$ and Auger coefficients below ~$10^{-29}$ $cm^6$ $s^{-1}$, beyond which rapid efficiency loss occurs due to severe carrier lifetime shortening and suppressed carrier collection. Thermal analyses indicate that device performance peaks near room temperature (~275 K), decreases afterwards and gradually saturates at higher temperatures, reflecting a dynamic balance between thermally induced carrier generation and recombination processes. Under varying illumination, the device exhibited linear $J_{SC}$ scaling (187.52 mA/W) and logarithmic $V_{OC}$ enhancement. PCE remains high across the examined range, showing no collapse at higher irradiances due to recombination or resistive effects, while maintaining substantial efficiency even under low-intensity conditions. By integrating optimized material, geometrical, and recombination parameters within physically realistic bounds, a peak simulated performance of $V_{OC}$ = 1.96 V, $J_{SC}$ = 18.75 mA $cm^{-2}$, FF = 86.44%, and PCE = 31.76% was achieved. These values represent theoretical performance projections under the current physically consistent modeling framework; practical devices may experience additional loss mechanisms—such as microstructural inhomogeneity, ion migration, and environmental degradation—that could moderate experimentally achieve efficiencies. Overall, this study establishes a physically grounded device-engineering framework for Cs$_2$AgBiBr$_6$ solar cells, demonstrating how tightly coupled optical–electrical modeling can convert material-level characteristics into quantitatively guided device optimization strategies. The framework established here is broadly applicable to emerging lead-free perovskite systems and offers a scalable pathway toward experimentally realizable, high-efficiency photovoltaic technologies.